\documentstyle[epsfig,preprint,aps,amsmath]{revtex}
\begin{document}
\tighten
\title{Near-threshold $\omega$ and $\phi$ meson productions in 
$pp$ collisions}
\author{K. Tsushima$^1$, and K. Nakayama$^{1,2}$}
\address{$^1$Department of Physics and Astronomy, University of Georgia, 
Athens, Georgia 30602, USA \\
$^2$Institut f\"{u}r Kernphysik, Forschungszentrum-J\"{u}lich,
D-52425, J\"{u}lich, Germany}
\maketitle


\begin{abstract}
Using a relativistic effective Lagrangian at the hadronic level, 
near-threshold $\omega$ and $\phi$ meson productions in proton proton 
($pp$) collisions, $p p \to p p \omega/\phi$, are studied within 
the distorted wave Born approximation. Both initial and final state 
$pp$ interactions are included. In addition to total cross section data, 
both $\omega$ and $\phi$ angular distribution data are  
used to constrain further the model parameters. For the 
$p p \to p p \omega$ reaction we consider two different 
possibilities: with and without the inclusion of nucleon resonances. 
The nucleon resonances are included in a way to be consistent with the 
$\pi^- p \to \omega n$ reaction. It is shown that the inclusion of 
nucleon resonances can describe the data better overall than 
without their inclusion. However, the SATURNE data in the range of 
excess energies $Q < 31$ MeV are still underestimated by 
about a factor of two. As for the $p p \to p p \phi$ reaction it is found 
that the presently limited available data from DISTO can be reproduced by 
four sets of values for the vector and tensor $\phi NN$ coupling constants. 
Further measurements of the energy dependence of the total cross 
section near threshold energies should help to constrain 
better the $\phi NN$ coupling constant.
\\ \\
PACS number(s): 25.40.Ve, 24.10.Jv, 24.30.-v, 13.75.Cs
\end{abstract}
%
%

\section{ Introduction}

Heavy meson production in nucleon nucleon ($NN$) 
collisions can, in principle, provide important information
about the short-range part of the $NN$ interaction~\cite{Moskal}.
For example, the $N N \to N N \omega/\phi$ reactions at their threshold   
energies probe distances between the two colliding nucleons  
of about $0.2$ fm~\cite{Kanzo_Cracow}. 
The distance corresponds to the ``overlapping region" 
of the two interacting nucleons, in contrast to the distance of about 
$0.5$ fm probed by much lighter pion production~\cite{Leepion}.
Therefore, investigation of such heavy meson production reactions 
should ultimately provide relevant information for testing QCD-based 
$NN$ interactions.

Recently, there has been considerable interest in the vector mesons  
$\omega$ and $\phi$, in connection with the OZI rule~\cite{OZI} violation,  
and in the strangeness content of the nucleon wave 
function~\cite{OZI_review,Lipkin,Ellis}. 
For example, the Crystal Barrel experiments at LEAR (CERN)~\cite{ppbar} 
found a strong violation of the OZI rule in the $\phi/\omega$ production rate 
in antiproton-proton ($\bar{p}p$) annihilation.
Furthermore, it was also found that the $\phi$ to $\omega$ production ratio 
in the $p p \to p p\, \omega/\phi$ reactions was enhanced by about 
an order of magnitude relative to the OZI prediction after correcting  
for the available phase space volume~\cite{DISTO}.
These findings may be interpreted as a considerable admixture of 
the $s\bar{s}$ configuration in the nucleon wave function.

Another item of interest in $\omega$ meson production processes    
is the so-called ``missing resonances" problem, 
where constituent quark models predict
more states than have been observed 
experimentally~\cite{Isgur,Capstick}.
This has been attributed to the possibility that many such missing 
resonances couple either weakly or not at all 
to the $\pi N$ channel, but may couple  
more strongly or exclusively to the $\omega N$ channel. 
Indeed, some theoretical studies of $\omega$ photoproduction 
were made~\cite{Zhao,Oh} inspired by this possibility.

Although, so far, no baryon resonances have been observed 
to decay into the $\omega N$ channel, some theoretical efforts 
have been made to estimate the coupling strengths of $\omega$ to 
the experimentally observed resonances~\cite{Riska,Mosel,Titov}.  
One of the motivations for such a study is the possibility 
to account for the observed 
enhancement of the low-mass dilepton pair production in heavy ion 
collisions~\cite{dilepton}. 
Alternatively, the downward shift of  
the $\rho$ and $ \omega$ meson masses (but a smaller   
shift for the $\phi$ meson mass) in the nuclear 
medium~\cite{Brown,Hatsuda,Asakawa,Koike,Klingl1} 
is also considered as a possible source of the observed enhancement.  
Indeed if the downward shift is large enough, $\omega$ meson is  
expected to form meson-nucleus bound 
states~\cite{Tsushima,Hayano,Klingl2}.
In any case a better understanding of the vector meson 
production mechanism in free space is 
a prerequisite to study such in-medium effects, and also to study the    
possible couplings of the $\omega$ meson to (missing) resonances.  
However, in spite of the pronounced interest in the 
vector mesons $\rho, \omega$ and $\phi$, so far there exist  
only a limited number of theoretical studies of the
near-threshold $p p \to p p \omega$~\cite{Alex,Kanzo_omega,Kaiser,Fuchs,TN} 
and $p p \to p p \phi$~\cite{Alex,Kanzo_phi,Titov2} reactions.
This situation also holds  for the near-threshold 
$p n \to d\, \omega/\phi$~\cite{Kanzo_d,Grishina} and
$p d \to ^3$He$\,\omega$~\cite{Wurzinger} reactions.

From the experimental side, apart from the old data at high excess 
energies~\cite{Flamino}, only the total cross section data from 
SATURNE~\cite{Hibou} were available until recently for 
$p p \to p p \omega$ in the near-threshold region with excess energies 
below $Q = 31$ MeV, where the excess energy $Q$ is defined as, 
$Q = \sqrt{s} - \sum_F m_F$ with $\sqrt{s}$ and $m_F$ being the 
total center-of-mass energy and masses of the particles in the 
final state, respectively.
There are also total cross section and angular 
distribution data at excess energy $Q = 319$ MeV from the 
DISTO Collaboration~\cite{DISTO}.
Recently the COSY-TOF Collaboration has measured the total 
cross section for $p p \to p p \omega$ at two excess energies,  
$Q = 92$ and $173$ MeV~\cite{COSY}. These fill in partly the energy 
gap between the SATURNE and DISTO data, 
and are critical in studying the energy dependence of the total cross section 
in the extended near-threshold region. 
In addition to the total cross sections, 
the COSY-TOF Collaboration has also measured the angular distribution of the 
$\omega$ meson produced at $Q = 173$ MeV. As has been pointed out 
in Ref.~\cite{Kanzo_omega}, the angular distribution plays a major 
role in disentangling different reaction mechanisms. These new data 
from the COSY-TOF Collaboration, together with earlier data~\cite{Hibou}, 
offer the opportunity to investigate the $p p \to p p \omega$ 
reaction more in detail than has been done previously. 
Thus, in the present study we focus on the near-threshold 
$p p \to p p \omega$ reaction in free space.
We study this reaction by considering two different possibilities: with and  
without the inclusion of nucleon resonances~\cite{TN}.  
The possibility of a large, off-shell $S_{11}(1535)$ resonance contribution
has been considered recently by the 
T\"{u}bingen group~\cite{Fuchs}. 

In the present work we also consider the $p p \to p p \phi$ reaction. 
The only data available for this reaction 
near-threshold energies are the reanalyzed total cross section 
and angular distribution from the DISTO Collaboration 
at $Q = 83$ MeV~\cite{DISTO}. These data are not sufficient 
to provide stringent constraints on theoretical models.
In studying the $p p \to p p \phi$ reaction 
we do not consider the possibility 
of nucleon resonances, because not enough 
data exist to either draw any meaningful conclusions about their role,   
or to fix new parameters associated with the resonances.
(Also note that no baryon resonances have been observed decaying 
into the $\phi N$ channel.)

To study the $p p \to p p \omega/\phi$ reactions, 
we use a relativistic effective Lagrangian at the 
hadronic level, where the reaction amplitude 
is calculated within the distorted wave Born approximation, including 
both the initial and final state $pp$ interactions  
(denoted by ISI and FSI, respectively). 
The ISI is implemented in the on-shell 
approximation~\cite{Kanzo_Cracow,TN,Kanzo_eta,Hanhart}, 
while the FSI is generated using the Bonn $NN$ potential model~\cite{Bonn}.
The finite width of the $\omega$, which is very important 
near threshold energies, is also included.
The $\omega p$ FSI is included only via the pole diagrams 
(s-channel processes). Many of the cut-off parameters and coupling 
constants necessary for the present study have already been fixed  
from other reactions in previous 
studies~\cite{Kanzo_Cracow,Kanzo_omega,Kanzo_phi,Kanzo_eta}.
It turns out that the $p p \to p p \omega$ reaction is apparently 
described better with the inclusion of nucleon resonances. 
However, in order to draw a definite conclusion we need 
more data for exclusive observables in the energy region above, but close to 
$Q = 30$ MeV. This is because there is no established method to  
remove the multi-pion background associated with the $\omega$-meson width
from the raw data in order to extract the cross sections. 
The finite $\omega$ width is very important for 
energies $Q < 30$ MeV and can possibly make the extracted data 
strongly dependent on the model used in the analysis~\cite{Hibou}. 
As for the $p p \to p p \phi$ reaction, we definitely need more data 
to constrain the model parameters.

This article is organized as follows.
In Section II the general structure of the reaction amplitude 
in the present approach is explained. 
In Section III we discuss the $p p \to p p \omega$ 
reaction without the inclusion of nucleon resonances;  
in Section IV we discuss the reaction  
with the inclusion of nucleon resonances. 
Section V treats the $p p \to p p \phi$ reaction.      
The results are discussed and summarized in Section VI.

\section{Structure of reaction amplitude}

In this section we describe the structure of the reaction amplitude     
for the $N N \to N N V\,(V=\omega,\phi)$ reaction following 
Ref.~\cite{Kanzo_eta}, to make this article self-contained.

\begin{figure}
\begin{center}
\vspace*{1cm}
\epsfig{file=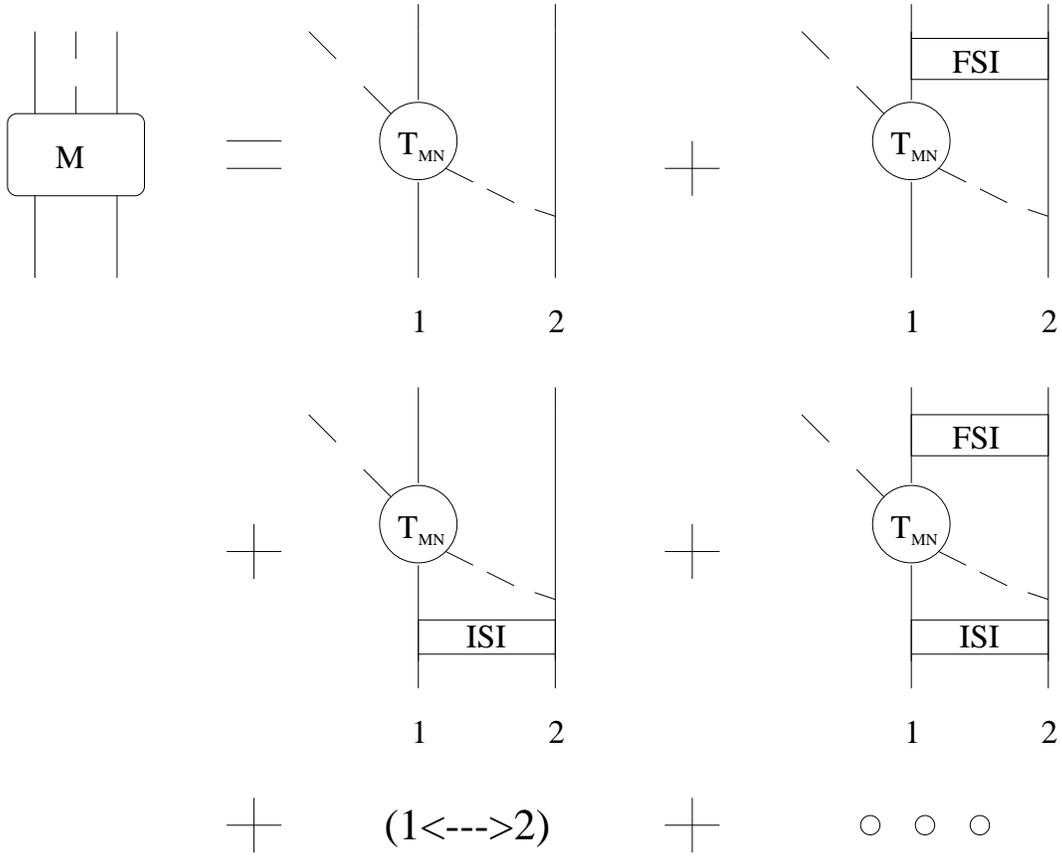,width=14cm}
\vspace{1cm}
\caption{Decomposition of the reaction amplitude for 
the $N N \to N N V \, (V=\omega,\phi)$ 
reaction in the present work.
$T_{MN}$ denotes the meson-nucleon $T$-matrix. 
ISI and FSI stand for the initial and final state 
$NN$ interactions, respectively.}
\label{fig_amplitude}
\end{center}
\end{figure}

In Fig.~\ref{fig_amplitude} we show a decomposition of the reaction 
amplitude for the $N N \to N N V\,(V=\omega,\phi)$ reactions. 
The  reaction amplitude is calculated in
the distorted wave Born approximation using a relativistic meson 
exchange model.
We begin by considering the meson-nucleon 
($MN$) and $NN$ interactions as the building blocks. 
Then, we consider all possible combinations of 
these building blocks in a topologically distinct way, with 
two nucleons in the initial state, and two nucleons plus a
meson in the final state. Here diagrams leading to double counting, 
e.g., those contributing to mass and vertex renormalizations 
must be excluded, since we use the physical masses and coupling constants.
The ellipsis in Fig.~\ref{fig_amplitude} indicates those diagrams that are 
more involved, or higher orders, which are not included in this work. 
In particular, we neglect the $MN$ FSI, 
which otherwise would be generated by solving the three-body 
Faddeev equation. 

In order to make use of the available potential models
of $NN$ scattering, we perform the calculation within a
three-dimensional formulation which is
obtained from the Bethe-Salpeter equation by maintaining the relativistic
unitarity and Lorentz covariance of the resulting amplitude. 
We follow the procedure of Blankenbecler and Sugar~\cite{Blank} as 
adopted in the Bonn $NN$ potential model~\cite{Bonn}. 
The vector meson production amplitude, $M$, may be written as
\begin{equation}
M = (1 + T^{(-)\dagger}_f i G^{(-)*}_f) 
(\epsilon^* \cdot J) (1 + i G^{(+)}_i T^{(+)}_i) \ ,
\label{ampl0}
\end{equation}
where $T_{(i,f)}$ stands for the $NN$ $T$-matrix 
in the initial($i$)/final($f$) state, 
$G_{(i,f)}$ is the three-dimensional
Blankenbecler-Sugar (BBS) propagator, and $\epsilon^*$ is the 
polarization vector of the vector meson produced.
The superscript $\pm$ in $T_{(i,f)}$ and $G_{(i,f)}$  
indicates the boundary conditions, $(-)$ for incoming and
$(+)$ for outgoing waves.
The production current $J^\mu$, which is 
the $MN$ $T$-matrix with one of the meson legs 
attached to a nucleon (first diagram on the r.h.s. in 
Fig.~\ref{fig_amplitude}) is defined by
\begin{equation}
J^\mu = \sum_{M'} \left[T_{(MN\leftarrow M'N)}\right]_1iP_{M'}
\left[\Gamma_{M'NN}^\mu\right]_2  +  (1 \leftrightarrow 2) \ ,
\label{curr0}
\end{equation}
where $T_{(MN\leftarrow M'N)}$ stands for the $MN$ $T$-matrix 
describing the transition $M'N \rightarrow MN$, and 
$\Gamma_{M'NN}^\mu$ and $P_{M'}$ are respectively the $M'NN$ vertex and the
corresponding meson propagator.
The subscripts 1 and 2 denote the two interacting
nucleons 1 and 2. The summation is over the intermediate mesons $M'$.
We note that if the four dimensional full two-nucleon propagator 
is used, the reaction amplitude given by Eq.~(\ref{ampl0}) 
would have an additional term to avoid the double counting arising 
from the term involving $iG_i^{(+)}T_i^{+}$ and/or $iG_f^{(+)}T_f^{+}$ 
and the current $J^\mu$, since $J^\mu$ also   
contains meson-exchange $NN$ interactions.
This additional term vanishes when we use the reduced three dimensional 
propagator $G_{(i/f)}$ which restricts the energy of the propagating 
two nucleons to be on their respective mass shells. 

In the near-threshold energy region, the two nucleon energy in the
final state $f$ is very low and hence 
the $NN$ FSI amplitude, $T^{(-)\dagger}_f$ in Eq.~(\ref{ampl0}), 
can be calculated from a number of realistic $NN$ potential models. 
In the present work we use the Bonn $NN$ potential model~\cite{Bonn}.
This model is defined by a reduced three dimensional BBS version of
the Bethe-Salpeter equation, 
\begin{equation}
T = {\cal V} + {\cal V} i G T \ ,
\label{Tmat}
\end{equation}
where $G$ denotes the BBS two-nucleon propagator, 
consistent with those appearing 
in Eq.~(\ref{ampl0}). (Note that the factor $-i$ difference in the 
definitions of ${\cal V}$ and $T$ from those in Ref.~\cite{Bonn}.)

For the $N N \to N N \omega/\phi$ reactions  
the $NN$ initial state interaction (ISI) amplitude, $T^{(+)}_i$
in Eq.~(\ref{ampl0}), must be calculated
at incident kinetic beam energies above $1.89$ and $2.59$ GeV, respectively.
There exists no accurate $NN$ interaction model with which one can perform 
calculations reliably at such high incident beam energies.
In the present work we follow Ref.~\cite{Hanhart},   
and make the on-shell approximation to evaluate the ISI contribution,  
which was also applied in the study of the 
$N N \to N N \eta$ reaction~\cite{Kanzo_eta}.
This amounts to keeping only the $\delta-$function part of the
Green function $G_i$ in evaluating the loop integral
involving $i G^{(+)}_i T^{(+)}_i$ in Eq.~(\ref{ampl0}).
The required on-shell $NN$ ISI amplitude is calculated from 
Ref.~\cite{CNS}.
As discussed in Ref.~\cite{Hanhart}, 
this is a reasonable approximation to the full $NN$ ISI. 
In this approximation the basic effect of the $NN$ ISI is to reduce 
the magnitude of the meson production cross 
section~\cite{Kanzo_eta,Lee}. 
In fact, it is easy to see that the angle-integrated production cross
section in each partial wave state 
$j$ is reduced by a factor $\lambda_j$~\cite{Hanhart}:
\begin{eqnarray}
\nonumber
\lambda_j & = & \left| \frac{1}{2} 
\left(\eta_j(p) e^{i2\delta_j(p)} + 1 \right) \right|^2
 \\ & = & \eta_{j}(p) \cos ^2(\delta_{j}(p)) + 
          \frac{1}{4}[1-\eta_{j}(p)]^2 \leq \frac{1}{4}[1+\eta_{j}(p)]^2 \ ,
\label{isieff}
\end{eqnarray}
where $\delta_j(p)$ and $\eta_j(p)$
denote respectively the phase shift and corresponding inelasticity, and  
$p$ is the relative momentum of the two nucleons in the initial state.
The $NN$ ISI has been considered fully by 
Batini$\acute{c}$ et al.~\cite{Lee} in the study of 
the $p p \to p p \eta$ reaction, whose threshold corresponds to 
an incident energy of about $1.25$ GeV. Their results support the 
on-shell approximation used in the present work. 
Quite recently, Baru et al.~\cite{Baru} have also investigated 
the effects of the $NN$ ISI in the $N N \to N N \eta$ reaction.
Although there are obviously (off-shell) effects which are absent in the 
on-shell approximation, the results of Ref.~\cite{Baru} also show that the 
bulk of the $NN$ ISI effect is accounted for in the on-shell 
approximation. In particular, we do not expect the off-shell 
effects of the ISI to change the conclusion of the present work.

Next, we consider the production current
$J^\mu$ defined by Eq.~(\ref{curr0}) based on meson exchange models. 
Following Refs.~\cite{Kanzo_phi,Kanzo_eta,Kanzo_etap},
we split the $MN$ $T$-matrix in Fig.~\ref{fig_amplitude} 
into the pole ($T^P_{MN}$) and non-pole ($T^{NP}_{MN}$) parts and
calculate the non-pole part in the Born approximation.
Then, the $MN$ $T$-matrix can be written as~\cite{Pearce}
\begin{equation}
T_{MN} = T^P_{MN} + T^{NP}_{MN} \ ,
\label{MNT}
\end{equation}
where
\begin{equation}
T^P_{MN} = \sum_B f_{MNB}^\dagger i g_B f_{MNB} \ ,
\label{MNTP}
\end{equation}
with $f_{MNB}$ and $g_B$ denoting the dressed
meson-nucleon-baryon ($MNB$) vertex and baryon
propagator, respectively. The summation is over the relevant baryons $B$. 
The non-pole part of the $T$-matrix is given by
\begin{equation}
T^{NP}_{MN} = V^{NP}_{MN} + V^{NP}_{MN}iGT^{NP}_{MN} \ ,
\label{MNTNP}
\end{equation}
where $V^{NP}_{MN}\equiv V_{MN} - V^P_{MN}$, with $V^P_{MN}$ 
denoting the pole part of the full $MN$ potential $V_{MN}$.
$V^P_{MN}$ is given by an equation analogous to Eq.~(\ref{MNTP}) 
with the dressed vertices and propagators replaced by the corresponding 
bare vertices and propagators. 
We neglect the second term of Eq.~(\ref{MNTNP}) and hence the
full $MN$ $T$-matrix in Eq.~(\ref{curr0}) is approximated
as $T_{MN} \cong T^P_{MN} + V^{NP}_{MN}$.

With the approximation described above, the resulting
current $J^\mu$ consists of baryonic and mesonic ($J_{mec}^\mu$) currents.
The baryonic current is further divided into
the nucleonic ($J_{nuc}^\mu$) and nucleon resonance ($J_{res}^\mu$) 
currents, so that the total current is written as 
\begin{equation}
J^\mu = J_{nuc}^\mu + J_{res}^\mu + J_{mec}^\mu \ .
\label{curr}
\end{equation}
The vector meson production currents are illustrated diagrammatically in 
Fig.~\ref{fig_current}, where $V$ stands for the  
$\omega$ or $\phi$ meson. 
Note that they are all Feynman diagrams and,
as such, they include both the positive- and negative-energy 
propagation of the intermediate state particles.
The nucleonic current is constructed consistently
with the $NN$ potential in Eq.~(\ref{Tmat}).
For the mesonic current, it turns out that we may 
consider only the $V\rho\pi$ ($V=\omega,\phi$) 
exchange-current contribution for both 
the $p p \to p p \omega/\phi$ reactions. 
The nucleon resonance current $J_{res}^\mu$, included only in 
the $p p \to p p \omega$ reaction, is explained in Section IV.   

Here, some general remarks on the meson production currents are in 
order~\cite{Kanzo_eta}. As a consequence of using a three-dimensional 
reduction of Bethe-Salpeter equation,  
the definition of the energy (time component) 
for the intermediate state particles in the production currents  
becomes ambiguous. In order to be consistent with the $NN$ interaction 
used in the present work, 
we follow the BBS three-dimensional reduction prescription:
(1) The energy of a virtual meson at the $MNN$ vertex is taken to be 
$q_0 = \varepsilon(p) - \varepsilon(p')$, where $\varepsilon(p) 
[\varepsilon(p')]$ is the energy of the nucleon before [after] 
the emission of the virtual meson, with 
$\varepsilon(p) \equiv \sqrt{p^2+m_N^2}$. 
(2) The energy of the intermediate state baryon B 
in the nucleonic and resonance 
currents is taken to be $p_0 = \omega(k) + \varepsilon(p')$ at the 
$B \to M + N$ vertex, while at the $N \to M + B$ vertex it is taken 
to be $p_0 = \varepsilon(p') - \omega(k)$, where $\omega(k)$ is the 
energy of the meson produced in the final state. 
The BBS reduction prevents three particle cuts which occur 
in a more exact calculation.

\section{$p p \to p p \omega$ without resonance}

\begin{figure}
\begin{center}
\vspace*{2cm}
\epsfig{file=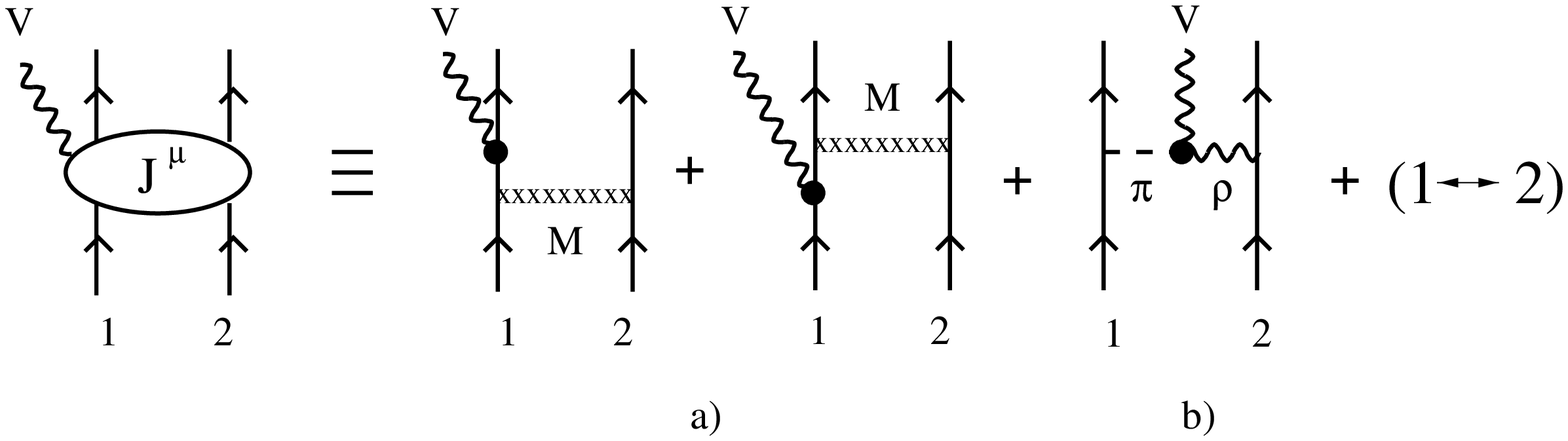,width=16cm,height=5cm}
\vspace{1cm}
\caption{Vector meson 
($\omega$ or $\phi$) production currents, $J^\mu$, included in the 
present study: a) nucleonic and resonance currents, b) mesonic current. 
$V = \omega$ or $\phi$ and M $= \pi,\eta,\rho,\omega,\sigma,a_0(=\delta)$. 
In the intermediate states of diagram a), negative-energy propagations 
for nucleon and resonances are included. 
}
\label{fig_current}
\end{center}
\end{figure}

In this Section, we consider the $p p \to p p \omega$ 
reaction without the inclusion of nucleon resonances.
Then, the total $\omega$-meson production 
current $J^\mu$ may be given by the sum of 
the nucleonic and $\omega\rho\pi$ meson-exchange currents,  
$J^\mu = J_{nuc}^\mu  + J_{mec}^\mu$, as shown in
Fig.~\ref{fig_current} ($V = \omega$). 

The nucleonic current $J_{nuc}^\mu$ is defined by 
\begin{equation}
J^\mu_{nuc} = \sum_{j=1,2}\left ( \Gamma^\mu_j iS_j U + U iS_j \Gamma^\mu_j 
\right ) \ ,
\label{nuc_cur}
\end{equation}
with $\Gamma^\mu_j$ denoting the $\omega NN$ vertex and 
$S_j$ the nucleon (Feynman)
propagator for the nucleon $j$. The summation is over the two interacting
nucleons, 1 and 2. $U$ stands for the meson-exchange $NN$ potential 
which is, in principle, identical to the driving potential 
${\cal V}$ used in the construction of the $NN$ interaction 
(Eq.~(\ref{Tmat})) , 
except that here meson retardation effects
are retained following the Feynman prescription. 

The $\omega$ production vertex $\omega NN$, $\Gamma^\mu_j$ 
in Eq.~(\ref{nuc_cur}), is obtained from the Lagrangian density, 
\begin{equation}
{\cal L}(x) =  - \bar\psi_N(x) \left( g_{\omega NN}
       [\gamma_\mu - \frac{\kappa_\omega}{2m_N}\sigma_{\mu\nu}\partial^\nu ] 
       \omega^\mu(x) \right) \psi_N(x) \ ,
\label{NNomega}
\end{equation}
where $m_N$, $\psi_N(x)$ and $\omega^\mu(x)$ stand for the nucleon mass,  
nucleon and $\omega$-meson fields, respectively. 
$g_{\omega NN}$ denotes the vector coupling constant and $\kappa_\omega 
\equiv f_{\omega NN}/g_{\omega NN}\, (g_{\omega NN}\ne 0)$, 
with $f_{\omega NN}$ the tensor coupling constant.
$J^\mu_{nuc}$ defined by Eq.~(\ref{nuc_cur})
is illustrated in Fig.~\ref{fig_current}a.

As in most meson-exchange models of hadronic interactions, each hadronic
vertex is accompanied by a form factor in order to account for
the composite or finite-size nature of 
the hadrons involved. 
Thus, the $\omega NN$ vertex obtained from the above Lagrangian
is multiplied by a form factor~\cite{Kanzo_phi,Kanzo_eta},  
\begin{equation}
F_{\omega NN}(p^2) = \frac{\Lambda_{N}^4} {\Lambda_{N}^4 + (p^2-m_N^2)^2}  \ ,
\label{formfactorN}
\end{equation}
where $p^2$ is the four-momentum squared of either the incoming or
outgoing off-shell nucleon. It is normalized to unity when the nucleon is
on its mass shell, $p^2 = m_N^2$.
Following Ref.~\cite{Kanzo_phi}, we adopt $g_{\omega NN} = 9.0$ in 
Eq.~(\ref{NNomega}). The vector to tensor coupling constant ratio, 
$\kappa_\omega$, is not well established; in fact the values quoted 
in literature are relatively small and vary in a range, 
$-0.16 \pm 0.01 \le \kappa_\omega \le +0.14 \pm 0.20$~\cite{Kanzo_omega}.
Therefore, we consider $\kappa_\omega$ and $\Lambda_N$, respectively in 
Eqs.~(\ref{NNomega}) and~(\ref{formfactorN}),  
as free parameters in the present work.

The $\omega\rho\pi$ vertex for $\omega$ production in the meson-exchange 
current, $J^\mu_{mec}$ (Fig.~\ref{fig_current}b), is derived from the 
Lagrangian density, 
\begin{equation}
{\cal L}_{\omega\rho\pi}(x) = \frac{g_{\omega\rho\pi}} {\sqrt{m_\omega m_\rho}}
\varepsilon_{\alpha\beta\nu\mu} \partial^\alpha \vec{\rho}^{\,\beta}(x) \cdot
\partial^\nu \vec{\pi}(x) \omega^\mu(x) \ ,
\label{pirhoomega}
\end{equation}
where $\varepsilon_{\alpha\beta\nu\mu}$ is the totally antisymmetric 
Levi-Civita tensor with $\varepsilon_{0123}=-1$. The $\omega\rho\pi$ 
vertex obtained from the above Lagrangian is multiplied by a form factor, 
\begin{equation}
F_{\omega\rho\pi}(q_\rho^2, q_\pi^2) \equiv
F_\rho (q_\rho^2) F_\pi (q_\pi^2) = 
\left (\frac{\Lambda_\rho^2}
            {\Lambda_\rho^2 - q_\rho^2}\right) 
\left (\frac{\Lambda_\pi^2 - m_\pi^2} {\Lambda_\pi^2 - q_\pi^2}\right ) \ , 
\label{formfactorM}
\end{equation}
where $\Lambda_\omega \equiv \Lambda_\rho = \Lambda_\pi$ 
is assumed~\cite{Kanzo_phi}.
It is normalized to unity at $q_\rho^2 = 0$ and $q_\pi^2 = m_\pi^2$, 
consistent with the kinematics at which the coupling constant 
$g_{\omega\rho\pi}$ is extracted. 

The meson-exchange current is given by
\begin{equation}
J^\mu_{mec} = [\Gamma^\alpha_{\rho NN}(q_\rho)]_1 iD_{\alpha\beta}(q_\rho)
              \Gamma^{\beta\mu}_{\omega\rho\pi}(q_\rho, q_\pi, k_\omega)
              i\Delta(q_\pi) [\Gamma_{\pi NN}(q_\pi)]_2   +  
(1\leftrightarrow 2) \ ,
\label{mec_cur}
\end{equation}
where $D_{\alpha\beta}(q_\rho)$ and $\Delta(q_\pi)$ stand for the $\rho$- and
$\pi$-meson (Feynman) propagators, respectively. The vertices $\Gamma$
involved are self-explanatory. 
The coupling constant, $g_{\omega\rho\pi}=10.0$ in Eq.~(\ref{pirhoomega}),
has been fixed from a systematic study of pseudoscalar and vector 
meson radiative decays combined with the vector meson dominance 
assumption~\cite{Kanzo_phi,Durso}.
Its sign is fixed from a study of pion photoproduction in the 
$1$ GeV energy region~\cite{Garcilazo}.
The $\rho NN$ and $\pi NN$ vertices in Eq.~(\ref{mec_cur}) are consistent  
with those in the Bonn-B $NN$ potential~\cite{Bonn}, except that here 
we use the pseudovector-coupling, 
which is consistent with the chiral constraints in the 
lowest order~\cite{Ulf}, instead of the pseudoscalar-coupling 
for the $\pi NN$ vertex. In addition, the cut-off parameter, 
$\Lambda_{\pi NN} = 1300$ MeV, is adopted at the $\pi NN$ vertex. 
(One could use the pseudoscalar-coupling, or even the admixture of the 
pseudoscalar- and the pseudovector-couplings~\cite{Kondratyuk}. 
Note that these two couplings entail  
different orders in chiral counting~\cite{Ulf}.
A test calculation by the use of the pseudoscalar-coupling with the 
same model parameters, turned out to give an enhancement of the total 
cross section by about a factor of ten near the threshold, and the
enhancement became far larger as the excess energy Q increases.)
We are, then, left with the cut-off parameter $\Lambda_\rho = \Lambda_\pi$ 
in Eq.~(\ref{formfactorM}) which will be treated as a free 
parameter in the present work.

Next, we explain how the model parameters 
$\kappa_\omega, \Lambda_N,$ and $\Lambda_\rho = \Lambda_\pi$ 
are determined in the present approach.
In Ref.~\cite{Kanzo_omega} it was pointed out that the angular 
distribution of the emitted $\omega$ mesons is a sensitive 
quantity for determining the {\it absolute amount} of 
nucleonic as well as mesonic current contributions 
in addition to the relative sign of the two amplitudes.
This was also demonstrated in Refs.~\cite{TN,JJHF}. 
(The value quoted in Ref.~\cite{JJHF} should read $g_{\omega NN} = + 9.0$.)
Furthermore, the shape of the $\omega$ angular 
distribution is particularly sensitive to the value of $\kappa_\omega$, 
the tensor to vector coupling ratio~\cite{TN,JJHF}.
Thus, we can make use of both the $\omega$ angular distribution 
and total cross section data 
from COSY-TOF~\cite{COSY} at $Q = 173$ MeV 
to fix these three parameters. 
We obtain a reasonable fit to the data 
with the values, $\Lambda_\rho = \Lambda_\pi = 1000$ MeV, 
$\Lambda_N = 1190$ MeV and $\kappa_\omega \simeq -2.0$.
Table~\ref{tab_nores} summarizes all the parameter values within the 
approach in this Section.
Here, it should be mentioned that, at $Q = 173$ MeV, the energy involved 
in the final $pp$ subsystem extends beyond the pion threshold.
The $NN$ FSI used in the present work has been developed to fit 
the phase-shifts only up to the pion production threshold. 
Therefore, strictly, we are beyond the applicability of this 
interaction. However, the final $pp$ energy involved is not large 
enough to introduce any significant deviation.

\begin{figure}
\begin{center}
\epsfig{file=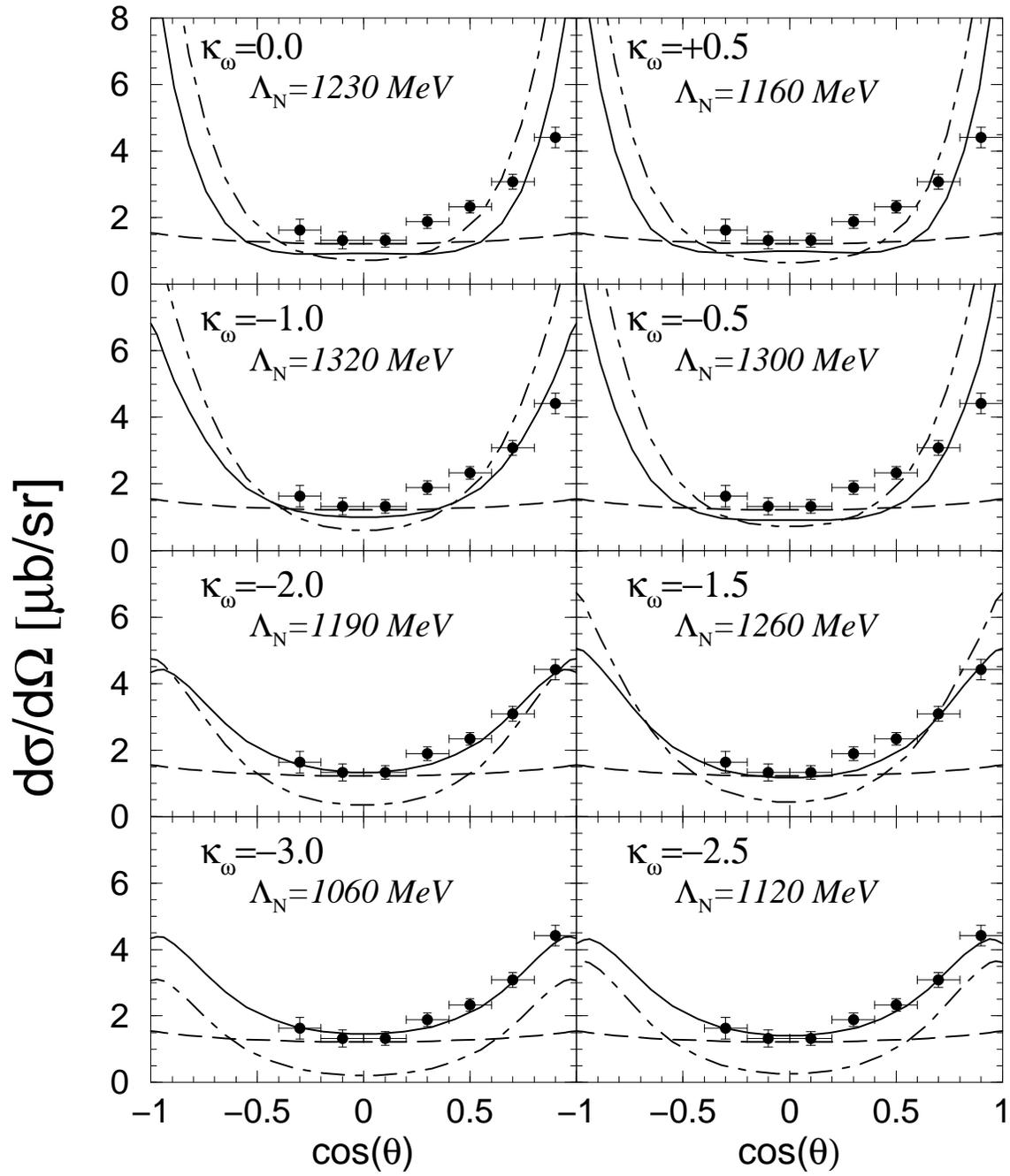,width=18cm,angle=-90}
\caption{$\kappa_\omega$ dependence of the $\omega$ angular 
distribution at excess energy $Q = 173$ MeV, 
without the inclusion of nucleon resonances. 
The (dashed, dot-dashed, solid) lines show the 
(mesonic, nucleonic, total) contributions, respectively.
The dots denote data from COSY-TOF~\protect\cite{COSY}.
The cut-off parameter $\Lambda_N$ in the form factor at  
the $\omega$ production vertex $\omega NN$ is fitted to 
the total cross section of $30.8 \mu b$ at excess energy  
$Q = 173$ MeV~\protect\cite{COSY} for each panel.
}
\label{fig_omangall}
\end{center}
\end{figure}

The $\kappa_\omega$ dependence of the $\omega$
angular distribution is illustrated in Fig.~\ref{fig_omangall}.
Here, since the mesonic current contribution gives 
a flat angular distribution, and because 
we are interested in the $\kappa_\omega$ dependence of the shape, 
we have kept the mesonic current contribution unchanged 
($\Lambda_\rho = \Lambda_\pi  = 1000$ MeV) and 
varied both $\kappa_\omega$ and $\Lambda_N$.  
The latter parameter has to be varied in order to keep the 
total cross section fixed. 
The results show indeed the shape
is sensitive to the value of $\kappa_\omega$.
Note in particular that values of $\kappa_\omega > -1$ are clearly unable 
to reproduce the data.
We note here that the value of the tensor coupling, 
$f_{\omega NN} = 0$, in the Bonn $NN$ potential used in the FSI 
in the present work, is quite different from the value of 
$f_{\omega NN} = \kappa_\omega g_{\omega NN} \simeq -2.0 \times 9 = -18$ 
used to reproduce the angular distribution data. 
However, the exchanged $\omega$ meson in the Bonn $NN$ potential model    
is associated with the virtual $\omega$ meson
and represents the isoscalar-vector quantum numbers exchanged,
and is not necessarily related to the physical $\omega$ meson.

\begin{figure}
\begin{center}
\noindent
\epsfig{file=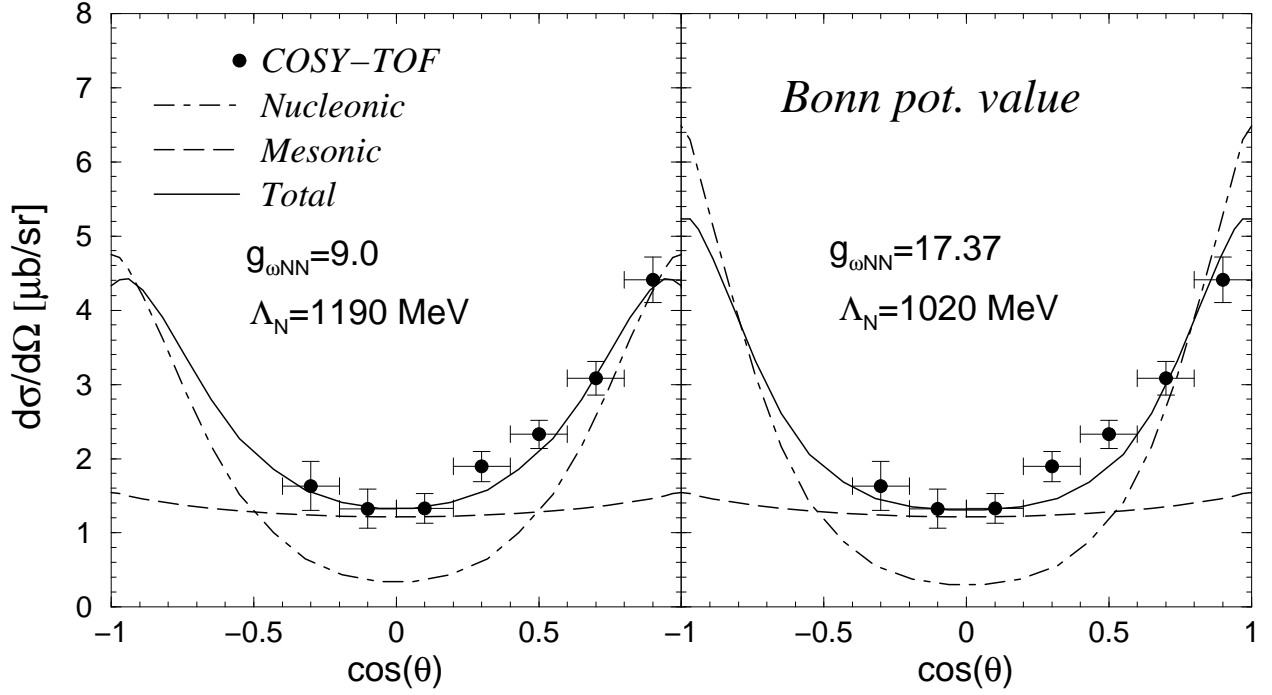,width=11cm,angle=-90}
\caption{$g_{\omega NN}$ dependence of the $\omega$ angular 
distribution at excess energy $Q = 173$ MeV. 
The left panel is one of the reasonable fits achieved with 
$g_{\omega NN} = 9.0$, while the right panel is the result obtained 
with the value $g_{\omega NN} = 17.37$, which is approximately the  
value used in the Bonn $NN$ potential model~\protect\cite{Bonn}. 
Both calculations use the value, $\kappa_\omega=-2.0$.
Also, see the caption of Fig.~\protect\ref{fig_omangall}.
}
\label{fig_omangBonn}
\end{center}
\end{figure}

Next, in order to see how the shape of the $\omega$ angular distribution 
is sensitive to the value of $g_{\omega NN}$, we keep   
the value, $\kappa_\omega = -2.0$ fixed, 
which reproduces the $\omega$ angular 
distribution very well, and calculate the $\omega$ angular 
distribution for different values of $g_{\omega NN}$. 
Here again the mesonic current contribution has been kept fixed. 
We show in Fig.~\ref{fig_omangBonn} (the right panel) 
the result obtained using the value, 
$(g_{\omega NN})^2/4\pi = (17.37)^2/4\pi = 24$, approximately the value 
used in the Bonn $NN$ potential, together with one of the reasonable 
fits (the left panel) obtained with $g_{\omega NN} = 9.0$.
Recall that the total cross section is normalized in both calculations.
The result obtained with $(g_{\omega NN})^2/4\pi = 24$ 
also gives a good fit to the data, 
where the change in the coupling constant $g_{\omega NN}$,  
$9.0 \to 17.37$, is compensated by the change in the cut-off parameter
$\Lambda_N$, $1190 \to 1020$ MeV.
This implies that, for a given mesonic current contribution, 
the shape of the $\omega$ angular distribution is not 
sensitive to the value of $g_{\omega NN}$, but is 
sensitive to the value of $\kappa_\omega$. This implies that 
the different momentum dependence introduced via 
the tensor coupling to the nucleonic current plays an important role to  
the shape of the $\omega$ angular distribution.

\begin{figure}
\begin{center}
\noindent
\epsfig{file=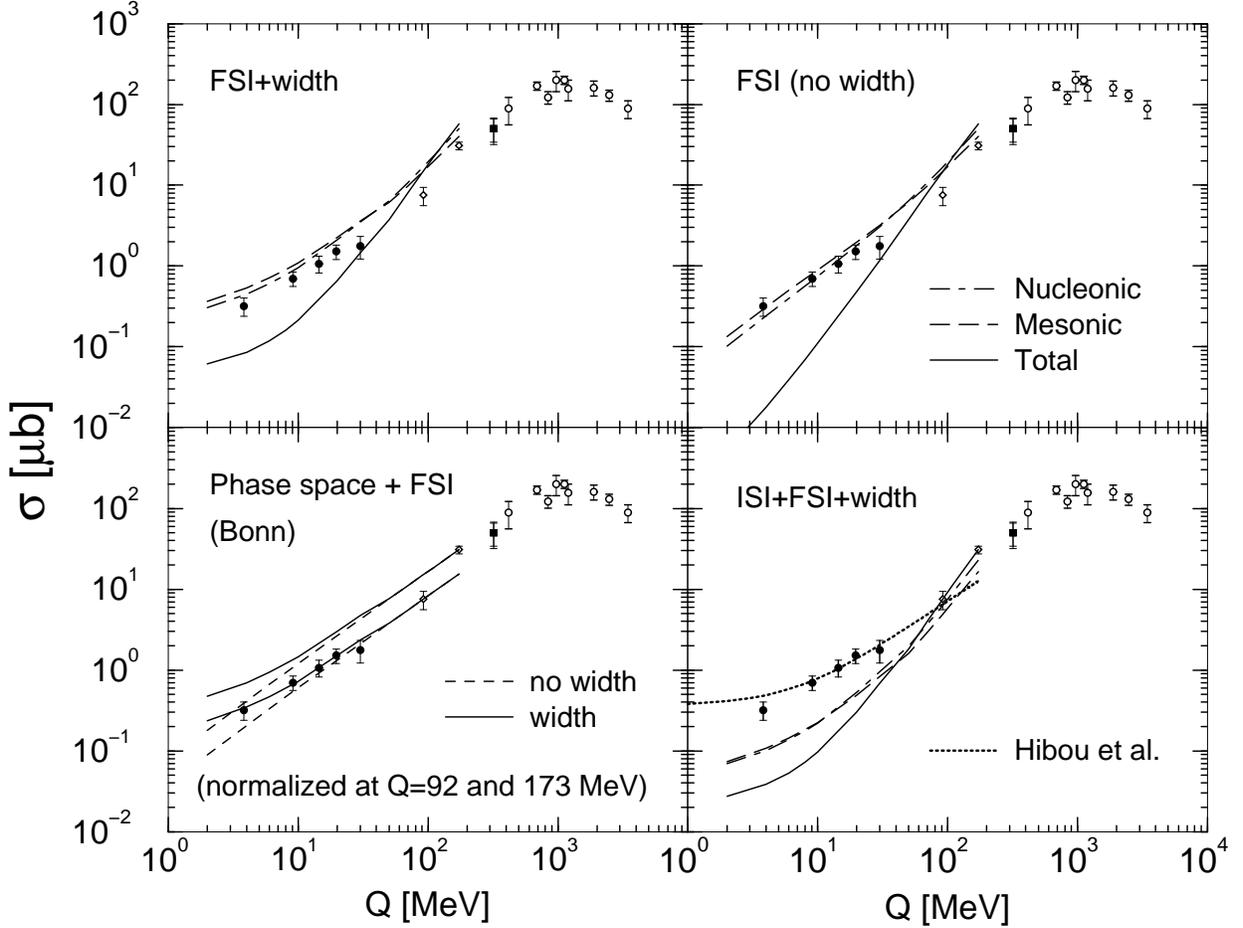,width=14cm,angle=-90}
\caption{Energy dependence of the total cross section for
the $p p \to p p \omega$ reaction without the inclusion of resonances.
``ISI", ``FSI" and ``width" stand for the $pp$ initial state interaction, 
$pp$ final state interaction, and effects of the $\omega$ width, 
respectively. The result with ``ISI+FSI+width" (the bottom-right panel) 
should be compared with the data, where those panels without 
any of the legends, ``ISI",``FSI" and ``width", 
imply that the corresponding 
effect is switched off from the full calculation (the bottom-right panel).
Data are from SATURNE~\protect\cite{Hibou} (dots), 
COSY-TOF~\protect\cite{COSY} (diamonds), 
Ref.~\protect\cite{Flamino} (circles) and DISTO~\protect\cite{DISTO} 
(a filled square), respectively. 
``Hibou et al." (the bottom-right panel)
stands for the result used in the analysis in
Ref.~\protect\cite{Hibou}.
In the bottom-left panel indicated by ``Phase space + FSI", 
the calculated energy dependences are normalized to the total cross 
section data from COSY-TOF~\protect\cite{COSY}
at either $Q = 92$ or $173$ MeV, where ``$(\epsilon^* \cdot J) =$ constant"
is used in Eq.~(\protect\ref{ampl0}), and thus, effects of the FSI 
generated by the Bonn NN potential model~\protect\cite{Bonn} enter.
}
\label{fig_noresxsec}
\end{center}
\end{figure}

With all the parameters fixed, we next study the energy dependence 
of the total cross section.
In Fig.~\ref{fig_noresxsec} we show the predicted energy dependence of 
the total cross section with various effects:
(1) effects of the finite $\omega$ width denoted by ``width",
(2) the initial and final state $pp$ interactions denoted by ``ISI" 
and ``FSI", respectively, and (3) using a constant matrix element 
denoted by ``Phase space", where ``$(\epsilon^* \cdot J) =$ constant"
is used in Eq.~(\ref{ampl0}).
The results with ``ISI+FSI+width" in Fig.~\ref{fig_noresxsec} 
(the bottom-right panel) include all the effects considered in the 
present section, and should be compared with the data. 
Recall that the present parameters are adjusted so as to reproduce 
both the total cross section of $30.8 \mu b$ and the $\omega$ angular 
distribution at $Q = 173$ MeV.
In Fig.~\ref{fig_noresxsec}, we also show the result of 
Ref.~\cite{Hibou} denoted by ``Hibou et al." used in the analysis 
of the SATURNE data.  
Obviously, the present result underestimates the SATURNE 
data~\cite{Hibou} to a large extent in the range of 
excess energies, $Q < 31$ MeV.
The reason for this discrepancy can be attributed to an 
overestimate of the mesonic current contribution 
as fixed at $Q = 173$ MeV. This results in a substantial reduction of 
the cross section close to threshold,
due to a much stronger destructive interference between
the nucleonic and mesonic current contributions as the excess
energy decreases.
In fact, the large overestimation of the mesonic current can be verified 
from the $\pi^- p \to \omega n$ reaction. With all the parameters 
fixed to reproduce the $p p \to p p \omega$ at $Q = 173$ MeV, 
we have looked at the model prediction of the energy dependence 
of the $\pi^- p \to \omega n$ total cross section. It turns out that 
the model largely overestimates the data from Ref.~\cite{piNdata} as the 
center-of-mass energy $W$ increases due to the rapidly increasing 
$\omega\rho\pi$ exchange current contribution. We were unable to 
reproduce both the $p p \to p p \omega$ data at $Q = 173$ MeV 
and the energy dependence of the $\pi^- p \to \omega n$ total cross section
within our approach which considers only the nucleonic and mesonic currents. 

In an attempt to improve the agreement in the present approach, 
we have investigated the effect of different form factors 
for the vector and tensor couplings at the $\omega$ production 
vertex in the nucleonic current. Different momentum dependences of the 
vector ($F_1$) and the tensor ($F_2$) strong form factors at 
the $\omega NN$ meson production vertex are 
quite possible. In particular, recent experimental results from the Jefferson 
Lab~\cite{Jlab} for the ratio of the proton electric and 
magnetic form factors, $\mu_p G_{Ep}(Q^2)/G_{Mp}(Q^2)$ 
($\mu_p$: proton magnetic moment), 
show a linear decrease as the four-momentum 
transfer squared ($Q^2$) increases, namely, 
the vector ($F_1$) to the tensor ($F_2$) electromagnetic
ratio is, $F_2(Q^2)/F_1(Q^2) \sim 1/Q$~\cite{Miller},
which shows that $F_1$ and $F_2$ have different momentum dependences. 
(The perturbative QCD predicts  
$F_2(Q^2)/F_1(Q^2) \sim 1/Q^2$~\cite{Brodsky}.) 
Using different cut-off values for the vector and tensor form factors,
respectively, $\Lambda_{Nv}=1300$ MeV and $\Lambda_{Nt}=1600$ MeV, 
and using the same functional form of Eq.~(\ref{formfactorN}), 
we have recalculated the energy dependence of the total cross section. 
Note that these parameters together with other parameters, 
$\Lambda_\rho = \Lambda_\pi = 1450$ MeV and $\kappa_\omega = -0.5$, 
can reproduce the $\omega$ angular distribution 
data at $Q=173$ MeV reasonably well. 
However, the predicted cross section is enhanced by only 
$10 \sim 20$ \% at near threshold energies, 
and still underestimates substantially  
all the SATURNE data points~\cite{Hibou}.
We have also considered a possible contribution of the 
$\omega\sigma\sigma$ mesonic current by 
assigning a reasonable range of values for the coupling constants 
and cut-off parameters associated with this current. 
But this also gives only a small contribution.
Thus, within the approach of considering only the contributions from the 
nucleonic and mesonic currents, 
it seems unlikely to be able to reproduce the measured 
energy dependence of the 
total cross section in the range of excess energies $Q \leq 173$ MeV.
Of course, we could fix the model parameters by fitting the total 
cross section at a lower excess energy point.  
However, the model would then overestimate the total cross sections near 
$Q = 173$ MeV substantially, and also it would be difficult to reproduce 
the $\omega$ angular distribution at $Q = 173$ MeV.

Apart from the difficulties mentioned above, we also note that the rather 
large (negative) value of $\kappa_\omega = -2.0$ required to 
reproduce the $\omega$ angular distribution, is not easily reconciled 
with other nuclear processes. For example, such a large value of 
$\kappa_\omega$ leads to a rather strong $NN$ (isoscalar) tensor 
force. This will affect the $NN$ tensor force, given primarily 
by the $\pi$ and $\rho$ meson exchange, in such a way that it 
becomes extremely difficult to describe NN scattering 
and deuteron properties. 
In fact, a rough calculation~\cite{Johann} shows that it is 
nearly impossible to describe the $NN$ phase-shift data 
with such a strong tensor coupling 
($f_{\omega NN} = \kappa_\omega g_{\omega NN} \simeq -18$). 

Thus, although there is currently no definite experimental
evidence for the $\omega$ meson to couple to any nucleon resonance,  
it would be natural to expect 
that some resonance currents give contributions, 
since high nucleon incident energies are 
involved in the near-threshold $\omega$ production in $pp$ collisions.
The reduction of the mesonic current at high excess energies may then be
compensated by the nucleon resonance current contributions.
We study such a possibility in Section IV.

\section{$p p \to p p \omega$ with resonance}

In order to limit the number of resonances considered and thereby 
avoid the introduction of an excessive number of new parameters, 
we restricted the resonances to those:
(1) that appreciably decay to the $N + \gamma$ channel   
so that the vector meson dominance (VMD) assumption 
may be used to produce $\omega$,
(2) whose mass distributions confined around  
$(m_N + m_\omega)$ and therefore contribute maximally at near threshold
energies, (3) that can describe consistently 
the $\pi^- p \to \omega n$ reaction. In addition,  
to see if a dominant $S_{11}(1535)$ resonance contribution 
as reported in Ref.~\cite{Fuchs} is possible,  
and since many parameters associated with $S_{11}(1535)$ 
are under better control than those for other higher resonances, 
we also include this resonance in the present study.
As a result, we consider contributions from the   
following four nucleon resonances, 
$S(1535)(\frac{1}{2}^-) ****, 
P(1710)(\frac{1}{2}^+) ***, 
D(1700)(\frac{3}{2}^-) ***$ and 
$P(1720)(\frac{3}{2}^+) ****$,  
where we list the spin, parity and status of the corresponding 
resonances explicitly~\cite{PDG}.
The construction of the resonance current, and the 
associated details are given in Ref.~\cite{Kanzo_eta}.

The resonance current $J_{res}^\mu$ contribution to the   
$p p \to p p \omega$ reaction arises from 
the spin-1/2 ($J_{1/2 res}^\mu$) and spin-3/2 ($J_{3/2 res}^\mu$) 
resonance currents in the present approach: 
\begin{equation}
J_{res}^\mu = J_{1/2 res}^\mu + J_{3/2 res}^\mu \ .
\label{res_cur}
\end{equation}
The spin-1/2 resonance current, in analogy to the nucleonic current, 
is defined by 
\begin{equation} 
J_{1/2 res}^\mu  =  
\sum_{j=1,2}\sum_{N^*}\left ( \Gamma_{\omega jN^*}^\mu 
iS_{N^*} U_{N^*} + \tilde{U}_{N^*} iS_{N^*} 
\Gamma_{\omega jN^*}^\mu \right ) 
\ .
\label{res12_cur}
\end{equation}
Here $\Gamma _{\omega jN^*}^\mu$ stands for the $\omega NN^*$ vertex
involving the nucleon $j$. 
$S_{N^*}(p) = ({\mbox{$ \not \! p$}} + m_{N^*}) / 
(p^2 - m_{N^*}^2 + im_{N^*}\Gamma_{N^*})$ 
is the $N^*$ resonance propagator, with 
$m_{N^*}$ and $\Gamma_{N^*}$ denoting the mass and width of the resonance,
respectively. The summation is over the two interacting nucleons, 
$j=1$ and $2$, and also over the spin-1/2 resonances, 
$N^* = S_{11}(1535)$ and $P_{11}(1710)$. 
In Eq.~(\ref{res12_cur}) $U_{N^*}$ ($\tilde{U}_{N^*}$) stands for the 
$NN\rightarrow NN^*$ ($NN^*\rightarrow NN$) 
meson-exchange transition potential, and is given by 
\begin{eqnarray}
U_{N^*} &=& \hspace{-2ex}
\sum_{M=\pi,\eta\,{\rm for}\,S_{11}(1535)\atop 
M=\pi,\eta,\sigma\,{\rm for}\,P_{11}(1710)}
\hspace{-2ex}
\Gamma_{MNN^*}(q) i\Delta_M(q^2)  \Gamma_{MNN}(q)
       + \sum_{M=\rho,\omega} \Gamma^\mu_{MNN^*}(q) iD_{\mu\nu (M)}(q)
\Gamma^\nu_{MNN}(q),   
\label{transpot12}
\end{eqnarray}
where $\Delta_M(q^2)$ and $D_{\mu\nu (M)}(q)$ are 
the (Feynman) propagators of the
exchanged pseudoscalar (scalar) and vector mesons, respectively.
$\Gamma_{MNN}(q)$ and $\Gamma^\mu_{MNN}(q)$ denote the pseudoscalar (scalar) 
and vector $MNN$ vertex, respectively. 
These vertices are taken consistently with the $NN$ potential
${\cal V}$ appearing in Eq.~(\ref{Tmat}), except for the type of 
coupling at the $\pi NN$ vertex and the $\omega NN$ coupling constant. 
(Recall that we use the pseudovector-coupling instead of the 
pseudoscalar-coupling at the $\pi NN$ vertex.) 
The exception also applies to the spin-3/2 resonance 
current, $J_{3/2 res}^\mu$.
An analogous expression to Eq.~(\ref{transpot12}) 
also holds for $\tilde U_{N^*}$.
In Eq.~(\ref{transpot12})  
we have an extra, $\sigma$-meson exchange in the 
$N P_{11}(1710) \leftrightarrow N N$ transition potential, 
since the $P_{11}(1710) \to N \pi \pi$ decay branch 
is relatively large~\cite{PDG}. 
Thus, we simulate these two pions conveniently by a $\sigma$-meson.

Following Refs.~\cite{Riska,Kanzo_eta,Benmer,Kanzo_inv}, 
the transition vertices $\Gamma_{MNN^*}$ and
$\Gamma^\mu_{MNN^*}$ in Eqs.~(\ref{res12_cur}) and~(\ref{transpot12}) 
for spin-1/2 resonances are obtained from the interaction 
Lagrangian densities, 
\begin{subequations}
\begin{eqnarray}
{\cal L}^{(\pm)}_{\eta NN^*}(x) & = & \mp g_{\eta NN^*}
\bar \psi_{N^*}(x) \left\{ \left[ i\lambda\Gamma^{(\pm)} + 
\left(\frac {1 - \lambda} {m_{N^*}\pm m_N}\right)\Gamma^{(\pm)}_\mu 
\partial^\mu \right] \eta(x) \right\} \psi_N(x) + h.c. \ , 
\label{NR12eta} \\
{\cal L}^{(\pm)}_{\pi NN^*}(x) & = & \mp g_{\pi NN^*}
\bar \psi_{N^*}(x) \left\{ \left[ i\lambda\Gamma^{(\pm)} + 
\left(\frac {1 - \lambda} 
{m_{N^*}\pm m_N}\right)\Gamma^{(\pm)}_\mu \partial^\mu\right] 
\vec{\tau} \cdot \vec{\pi}(x) \right\} \psi_N(x) + h.c. \  ,
\label{NR12pi} \\
{\cal L}_{\sigma NP_{11}}(x) & = & g_{\sigma NP_{11}}
\bar \psi_{P_{11}}(x) \sigma \psi_N(x) + h.c. \  , 
\label{NR12sigma} \\
{\cal L}^{(\pm)}_{\omega NN^*}(x) & = &
\left(\frac {g_{\omega NN^*}} {m_{N^*}+m_N}\right)
\bar \psi_{N^*}(x) \left\{
\left[ \frac{\Gamma_\mu^{( \mp )}\partial^2}{m_{N^*}+m_N} 
+ \Gamma^{( \mp )}\left( - i\partial_\mu + 
\kappa_\omega\sigma_{\mu\nu} \partial^\nu \right) \right]
\omega^\mu(x) \right\} \psi_N(x) \nonumber \\ 
& + & h.c.  \ ,
\label{NR12omega} \\
{\cal L}^{(\pm)}_{\rho NN^*}(x) & = & 
\left(\frac {g_{\rho NN^*}} {m_{N^*}+m_N}\right)
\bar \psi_{N^*}(x) \left\{
\left[ \frac{\Gamma_\mu^{( \mp )}\partial^2}{m_{N^*}+m_N} 
+ \Gamma^{( \mp )}\left( -i\partial_\mu + 
\kappa_\rho\sigma_{\mu\nu} \partial^\nu \right) \right]
\vec \tau \cdot \vec \rho^{\,\mu }(x) \right\}\psi_N(x) 
\nonumber \\
& + & h.c.  \ ,
\label{NR12rho}
\end{eqnarray}
\label{NR12M}
\end{subequations}
where $\vec{\pi}(x)$, $\omega^\mu(x)$, $\vec\rho^{\,\mu}(x)$ and 
$\psi_{N^*}(x)$ denote the $\pi$, $\omega$, $\rho$ and spin-1/2 
nucleon resonance fields, respectively. The upper
and lower signs refer to the even(+) and odd(-) parity 
resonances, respectively. The operators $\Gamma^{(\pm)}$ and 
$\Gamma^{(\pm)}_\mu$ in Eqs.~(\ref{NR12eta})~-~(\ref{NR12rho}) 
are defined by
\begin{equation}
\left(\Gamma^{(+)}, \Gamma^{(-)}, \Gamma^{(+)}_\mu, \Gamma^{(-)}_\mu\right) 
= \left(\gamma_5, 1, \gamma_5 \gamma_\mu, \gamma_\mu\right) \ .
\label{NRpsa}
\end{equation}
The parameter $\lambda$ in Eqs.~(\ref{NR12eta}) and~(\ref{NR12pi}) 
controls the admixture of the two types of couplings: 
pseudoscalar (ps-coupling: $\lambda=1$) and pseudovector 
(pv-coupling: $\lambda=0$) 
for an even parity resonance and, scalar ($\lambda=1$) and vector 
($\lambda=0$) for an odd parity resonance, where both choices of
the parameter $\lambda$ give equivalent results when baryons are 
on their mass shells. In this work we take $\lambda=0$. 
Note that in principle we should not allow only 
the pure $\Gamma^{(\mp)}_\mu$ coupling in Eqs.~(\ref{NR12omega}) 
and~(\ref{NR12rho}), because unlike the $VNN$ vertex 
($V=$vector meson), this coupling alone at the $VNN^*$ vertex prevents 
us from estimating its strength using the VMD, since it violates 
gauge invariance. In the present work, we use a more general 
gauge invariant Lagrangian density as used in Ref.~\cite{Kanzo_inv}  
based on Ref.~\cite{Riska}.

Similar to the case of spin-1/2 resonances, the spin-3/2 resonance 
current is defined by 
\begin{equation}
J_{3/2 res}^\mu  =  \sum_{j=1,2}\sum_{N^*}\left ( 
\Gamma^{\mu\alpha}_{\omega jN^*} iS_{\alpha\beta (N^*)} U^\beta_{N^*} + 
\tilde U^\alpha_{N^*} iS_{\alpha\beta (N^*)} 
\Gamma^{\beta\mu}_{\omega jN^*} \right ) \ .
\label{res32_cur}
\end{equation}
Here $\Gamma^{\beta\mu}_{\omega jN^*}$ stands for the 
$\omega NN^*$ vertex function involving
the nucleon $j$. $S_{\alpha\beta (N^*)}(p) = ({\mbox{$ \not \! p$}} + m_{N^*})
\left\{- g_{\alpha\beta} + \gamma_\alpha\gamma_\beta/3
+ (\gamma_\alpha p_\beta - p_\alpha\gamma_\beta)/3m_{N^*} 
+ 2p_\alpha p_\beta / 3m^2_{N^*} \right\}
/ (p^2 - m_{N^*}^2 + im_{N^*}\Gamma_{N^*})$
is the spin-3/2 Rarita-Schwinger propagator. 
The summation is over the two interacting
nucleons, $j=1$ and $2$, and also over the spin-3/2 resonances,  
$N^* = D_{13}(1700)$ and $P_{13}(1720)$.  
In Eq.~(\ref{res32_cur}) $U^\alpha_{N^*}$
($\tilde U^\alpha_{N^*}$) stands for the 
$NN\rightarrow NN^*$ ($NN^*\rightarrow NN$) 
meson-exchange transition potential, and is given by
\begin{equation}
U^\alpha_{N^*} = \sum_{M=\pi, \eta} \Gamma^\alpha_{MNN^*}(q) 
i\Delta_M(q^2)  \Gamma_{MNN}(q)
+ \sum_{M=\rho, \omega} \Gamma^{\alpha\lambda}_{MNN^*}(q) 
iD_{\lambda\nu (M)}(q) \Gamma^\nu_{MNN}(q)   \ ,
\label{transpot32}
\end{equation}
where $\Gamma^\alpha_{MNN^*}(q)$ and 
$\Gamma^{\alpha\lambda}_{MNN^*}(q)$ denote the
pseudoscalar and vector $MNN^*$ vertices, respectively. 
An analogous expression to Eq.~(\ref{transpot32}) also holds for 
$\tilde{U}^\alpha_{N^*}$.

The $MNN^*$ vertices involving spin-3/2 nucleon resonances in
Eqs.~(\ref{res32_cur}) and~(\ref{transpot32}) are obtained from the 
Lagrangian densities~\cite{Kanzo_eta,Benmer},  
\begin{subequations}
\begin{eqnarray}
{\cal L}^{(\pm)}_{\eta NN^*}(x) & = & \left(\frac {g_{\eta NN^*}} 
{m_\eta}\right) \bar \psi_{N^*}^\mu(x) \Theta_{\mu\nu}(z) \Gamma^{(\mp)} 
\psi_N(x) \partial^\nu \eta(x) + h.c. \ ,
\label{NR32eta} \\
{\cal L}^{(\pm)}_{\pi NN^*}(x) & = & \left(\frac {g_{\pi NN^*}} {m_\pi}\right)
\bar \psi_{N^*}^\mu(x) \Theta_{\mu\nu}(z) \Gamma^{(\mp)} \vec{\tau} \psi_N(x)
\cdot \partial^\nu \vec{\pi}(x) + h.c. \ ,
\label{NR32pi} \\
{\cal L}^{(\pm)}_{\omega NN^*}(x) & = & 
\mp i \left(\frac {g_{\omega NN^*}^{(1)}} {2m_N}\right)
\bar \psi_{N^*}^\mu(x) \Theta_{\mu\nu}(z) \Gamma^{(\pm)}_\lambda 
\psi_N(x) \omega^{\lambda\nu}(x) \nonumber \\ 
& - &  \left(\frac {g_{\omega NN^*}^{(2)}} {4m_N^2}\right) \left( 
\partial_\lambda \bar \psi_{N^*}^\mu(x) 
\Theta_{\mu\nu}(z) \Gamma^{(\pm)} \psi_N(x) \right)
\omega^{\lambda\nu}(x)  + h.c. \  ,
\label{NR32omega} \\
{\cal L}^{(\pm)}_{\rho NN^*}(x) & = & 
\mp i \left(\frac {g_{\rho NN^*}^{(1)}} {2m_N}\right) \bar \psi_{N^*}^\mu(x) 
\Theta_{\mu\nu}(z) \Gamma^{(\pm)}_\lambda \vec{\tau} \psi_N(x) \cdot 
\vec{\rho}^{\,\lambda\nu}(x) \nonumber \\
& - &  \left(\frac {g_{\rho NN^*}^{(2)}} {4m_N^2}\right) \left( 
\partial_\lambda \bar \psi_{N^*}^\mu(x) 
\Theta_{\mu\nu}(z) \Gamma^{(\pm)} \vec{\tau} \psi_N(x) \right) \cdot
\vec{\rho}^{\,\lambda\nu}(x) + h.c. \  ,
\label{NR32rho} 
\end{eqnarray}
\label{NR32M}
\end{subequations}
where $\Theta_{\mu\nu}(z) \equiv g_{\mu\nu} - (z+1/2)\gamma_\mu \gamma_\nu$, 
and $\omega^{\lambda\nu}(x)\equiv \partial^\lambda\omega^\nu(x) 
- \partial^\nu\omega^\lambda(x)$ and
$\vec{\rho}^{\,\lambda\nu}(x)\equiv \partial^\lambda\vec{\rho}^{\,\nu}(x) 
- \partial^\nu\vec{\rho}^{\,\lambda}(x)$.
In order to reduce the number of parameters, 
we take $z=-1/2$ in the present work.

Following Ref.~\cite{Kanzo_eta}, the relevant coupling constants associated 
with the resonance currents are calculated utilizing the 
Particle data~\cite{PDG} whenever available; 
they are determined from the centroid values of the extracted partial 
decay widths (and masses) of the resonances. 
Those couplings involving vector mesons, are estimated
from the corresponding measured radiative decay width 
in conjunction with the VMD, although uncertainties in the data 
are large.
In order to reduce the number of free parameters, 
the ratio of the $VNN^*$ ($V=\rho,\omega$) coupling constants for the
spin-3/2 resonances, 
$D_{13}(1700)$ and $P_{13}(1720)$, have been fixed to be 
$g_{VNN^*}^{(1)}/g_{VNN^*}^{(2)}=-2.1$, the same ratio as that for   
$g_{\gamma NP_{33}(1232)}^{(1)}/g_{\gamma NP_{33}(1232)}^{(2)}=-2.1$, 
extracted from the ratio of 
$E2/M1 \cong -2.5\%$ determined from pion
photoproduction measurements~\cite{MAINZ}.

Following Refs.~\cite{Kanzo_eta,Kanzo_etap}, and 
in complete analogy to the nucleonic current,
we introduce the off-shell form factors at each vertex involved 
in resonance currents. We adopt the same form factor as given by 
Eq.~(\ref{formfactorN}), with $m_N$ replaced by $m_{N^*}$ at 
the $MNN^*$ vertex, in order to account for the
$N^*$ resonance being off-shell. The $MNN^*$ vertex, where the exchanged
meson is also off-shell, is multiplied by an extra form factor $F_M(q^2)$ in
order to account for the meson being off-shell 
(see Eqs.~(\ref{transpot12}) and~(\ref{transpot32})). 
The corresponding full form factor is, therefore, given by the product 
$F_N(p^2)F_M(q^2)$, where $M$ stands for the 
exchanged meson between the two interacting nucleons. 
The form factor $F_M(q^2)$ is taken consistently with the $NN$ potential 
${\cal V}$ in Eq.~(\ref{Tmat}); the only two differences are the normalization 
point of $F_{\rho, \omega}(q^2)$ and the cutoff parameter value of 
$F_\pi(q^2)$. Here, the form factor for vector mesons
$F_{\rho,\omega}(q^2)$ is normalized to unity at $q^2=0$ in accordance with 
the kinematics at which the coupling constant $g_{\rho,\omega NN^*}$ was 
extracted, i.e., 
$F_{\rho, \omega}(q^2) 
= (\Lambda_{\rho,\omega}^2 / (\Lambda_{\rho,\omega}^2 - q^2))^2$. 
For the pion form factor $F_\pi(q^2)$ we  
use the cutoff value $\Lambda_\pi=900$ MeV. 
To be consistent in this section, this value 
(rather than $\Lambda_\pi = 1300$ MeV) is also used in the form 
factor at the $\pi NN$ vertex appearing in the mesonic current constructed.  

Since we now include the resonance current, 
the free parameters in the nucleonic and mesonic currents 
in the previous Section have to be readjusted. 
In addition, we will consider the $\sigma N N^*$ coupling for 
$N^* = P_{11}(1710)$ as a free parameter in the present work.
In order to fix these free parameters in this Section, we use the 
$\pi^- p \to \omega n$ total cross section data from Ref.~\cite{piNdata}
(see Ref.~\cite{Mosel} for a discussion about the data), 
in addition to the $p p \to p p \omega$ 
total cross section and angular distribution data from the COSY-TOF 
Collaboration~\cite{COSY}.
We note that at the excess energy of $Q = 173$ MeV for the 
$p p \to p p \omega$ reaction,
the center-of-mass energy $W$ of the subsystem $\pi^- p \to \omega n$ 
appearing as a building block in the description of 
the $p p \to p p \omega$ reaction,
will reach a maximum value of $W \simeq 1.9$ GeV.
Thus, we fix the parameters so as to reproduce  
the measured energy dependence of the $\pi^- p \to \omega n$ 
total cross section data up to $W \simeq 1.9$ GeV.

\begin{figure}
\begin{center}
\epsfig{file=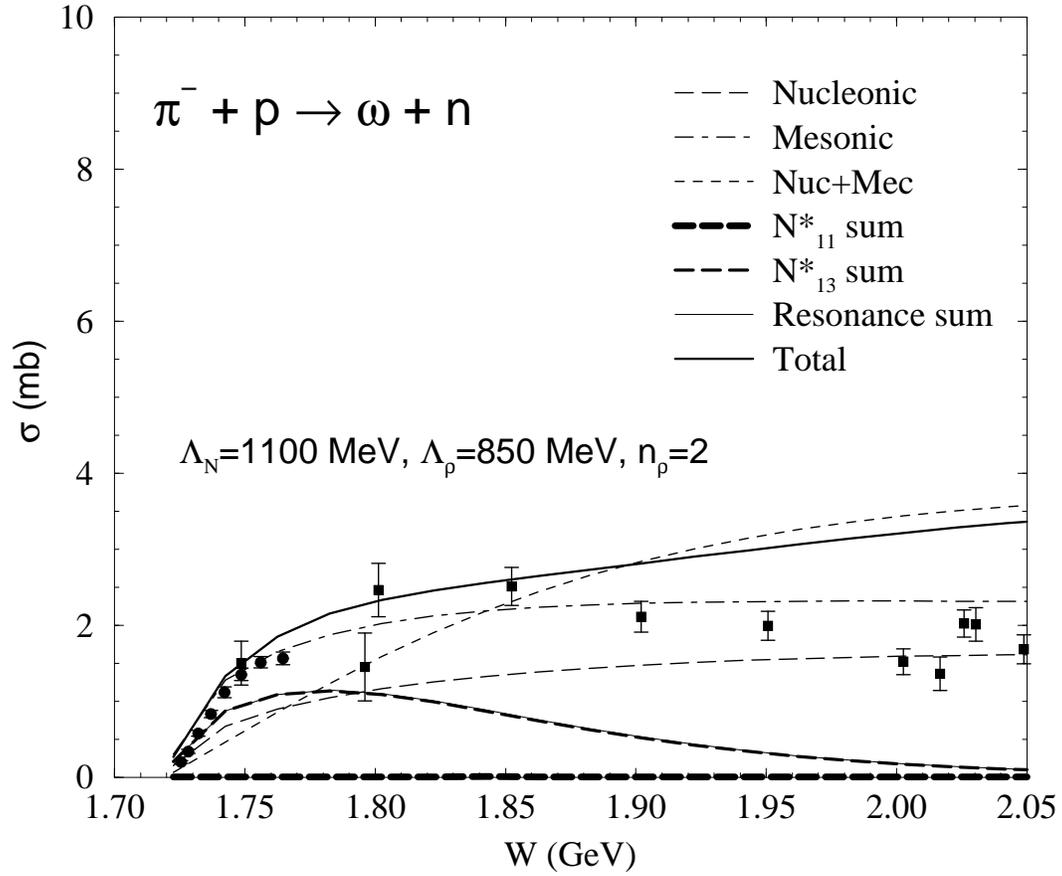,height=14cm,angle=-90}
\caption{Energy dependence of the total cross section for 
the $\pi^- + p \to \omega n$ reaction obtained with the 
preferred model parameter set.
Data are from Ref.~\protect\cite{piNdata}. 
Note that at an excess energy of $Q = 173$ MeV in the 
$p p \to p p \omega$ reaction, the maximum center-of-mass 
energy $W$ for this reaction reaches $\simeq 1.9$ GeV.
}
\label{fig_pinon}
\end{center}
\end{figure}

We show in Fig.~\ref{fig_pinon} the calculated energy 
dependence of the total cross section obtained with a selected 
parameter set. At lower energies $W$, $D_{13}(1700)$ and $P_{13}(1720)$ 
contributions are dominant, but neither $S_{11}(1535)$ nor $P_{11}(1710)$
give appreciable contributions for the $\pi^- p \to \omega n$ total 
cross section. Furthermore,  
many trial calculations show that, without including 
the resonances it is very difficult to reproduce the near-threshold 
behavior of the $\pi^- p \to \omega n$ total cross section 
up to energies $W \simeq 1.9$ GeV using a reasonable set of parameters.
However, with the inclusion of the resonances, we need a 
softer (stronger) form factor for the $\omega \rho \pi$ 
vertex to fit the overall 
energy dependence exhibited by the data. 
In particular, the part of the form factor which
accounts for the off-shell behavior of the exchanged 
$\rho$ meson requires a dipole form, 
\begin{equation}
F_{\omega\rho\pi}(q_\rho^2, q_\pi^2) \equiv 
F_\rho (q_\rho^2) F_\pi (q_\pi^2) = 
\left (\frac{\Lambda_\rho^2}
            {\Lambda_\rho^2 - q_\rho^2}\right)^{n_\rho} 
\left (\frac{\Lambda_\pi^2 - m_\pi^2} {\Lambda_\pi^2 - q_\pi^2}\right ) \ , 
\label{formfactorM2}
\end{equation}
with $n_\rho=2, \Lambda_\rho=850$ MeV and $\Lambda_\pi=1450$ MeV.
A cutoff parameter value of $\Lambda_N = 1100$ MeV has been also determined
at the $\omega NN$ meson production vertex.
A different form factor, $\exp (\beta q_\rho^2) \exp(-\alpha W^2)$, 
was introduced at the $\omega\rho\pi$ vertex in Ref.~\cite{Alex_omega} 
to overcome the difficulties in reproducing the data 
using the monopole form factor $F_\rho (q_\rho^2)$ 
in Eq.~(\ref{formfactorM2}). 

Next, using the $\omega$ angular distribution data from COSY-TOF~\cite{COSY}, 
we further fix parameters associated with the $P_{11}(1710)$ resonance, 
namely, the coupling constant $g_{\sigma N P_{11}}$ associated with the 
$\sigma$ exchange introduced effectively to simulate the 
observed decay channel, $P_{11}(1710) \to N + 2\pi$.
The value for the coupling constant $g_{\sigma NP_{11}}$ 
is adjusted to reproduce the $p p \to p p \omega$ total cross 
section of $30.8 \mu b$ at $Q = 173$ MeV.
Thus, in this procedure, the value obtained for $g_{\sigma NP_{11}}$ 
is not strictly related to the branching ratio for the $N + 2\pi$ 
channel; instead, the contribution from $P_{11}(1710)$ should be regarded 
as also taking into account the other possible resonance contributions 
not included explicitly in our model.

\begin{figure}
\begin{center}
\epsfig{file=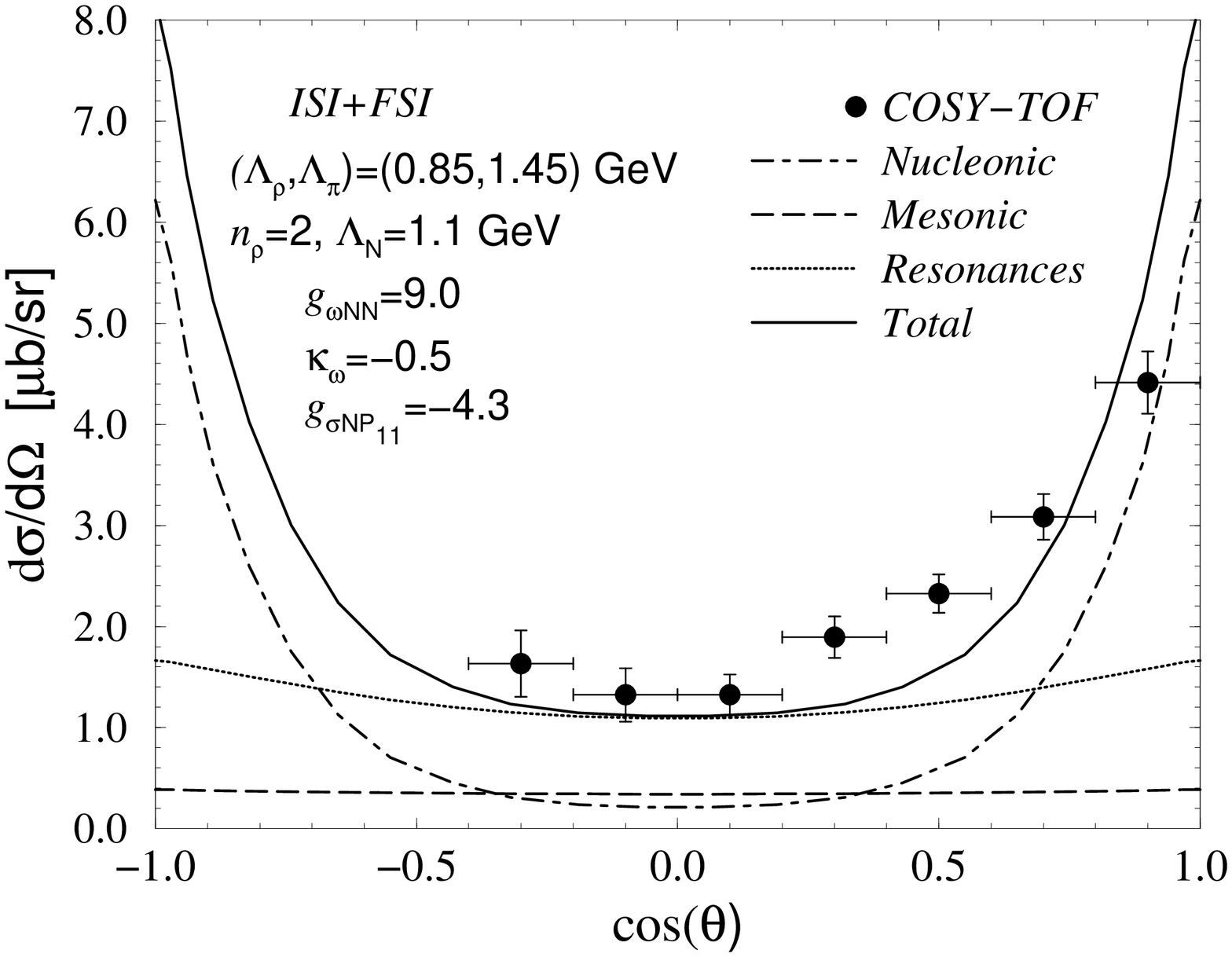,width=9cm,height=14cm,angle=-90}
\epsfig{file=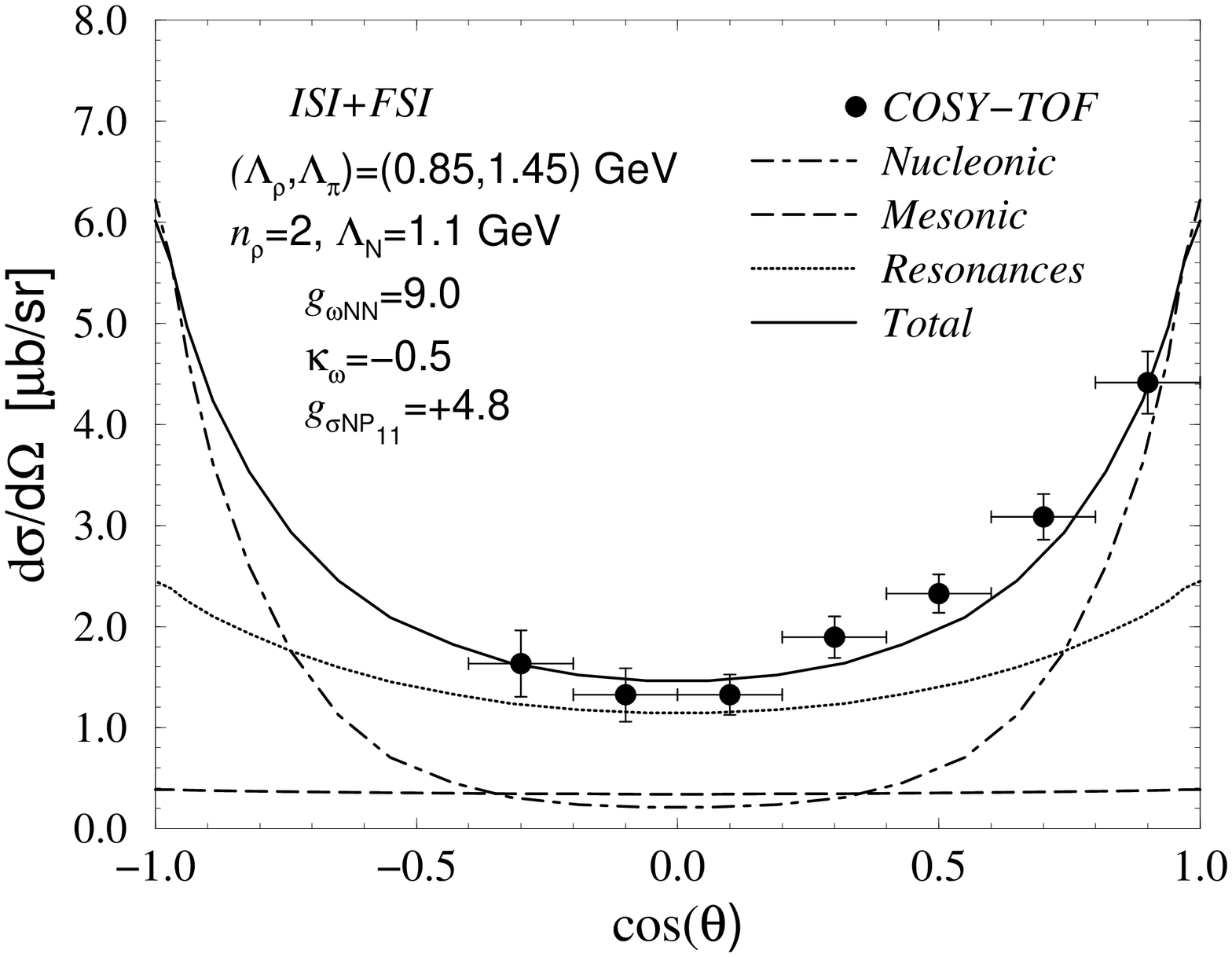,width=9cm,height=14cm,angle=-90}
\caption{Calculated $\omega$ angular distribution 
for the $p p \to p p \omega$ reaction at 
excess energy $Q = 173$ MeV.
The cut-off parameter $\Lambda_N$ in the $\omega NN$ form factor, 
$F_N(p^2) = \Lambda_N^4/[\Lambda_N^4+(p^2-m_N^2)]$  
[Eq.~(\protect\ref{formfactorN})],
is adjusted to the $\pi^- p \to \omega n$ reaction $\Lambda_N=1100$ MeV 
(Fig.~\protect\ref{fig_pinon}), and the coupling constant 
$g_{\sigma NP_{11}}$ is fitted to reproduce the total cross section 
of $30.8 \mu b$ at $Q = 173$ MeV.
The cut-off parameters and $n_\rho$ indicated in each panel enter at the 
$\omega\rho\pi$ vertex form factor:  
$F_{\pi\rho\omega}(q_\pi^2,q_\rho^2) 
\equiv F_\rho (q_\pi^2) \times F_\pi (q_\rho^2) = 
\left[{\Lambda_\rho^2}/(\Lambda_\rho^2 - q_\rho^2)\right]^{n_\rho} \times 
\left[(\Lambda_\pi^2 - m_\pi^2)/(\Lambda_\pi^2 - q_\pi^2)\right]$
[Eq.~(\protect\ref{formfactorM2})], 
with $n_\rho = 2$, $\Lambda_\rho = 850$ MeV and 
$\Lambda_\pi = 1450$ MeV, respectively.
Also, see the caption of Fig.~\protect\ref{fig_omangall}.
}
\label{fig_omangres}
\end{center}
\end{figure}

We show in Fig.~\ref{fig_omangres} the  
$\omega$ angular distribution calculated by fitting the coupling 
constant $g_{\sigma N P_{11}}$ to the total cross section of 
$30.8 \mu b$, together with a more reasonable value of $\kappa_\omega=-0.5$. 
Recall that with the value of $\kappa_\omega \simeq -2.0$ 
obtained in Section III, it would be very difficult to describe 
the $NN$ scattering data consistently~\cite{Johann}.
Two values for $g_{\sigma NP_{11}}$ are found to be able to 
reproduce the total cross section of $30.8 \mu b$ at $Q = 173$ MeV. 
The results for the $\omega$ angular distribution are shown 
in Fig.~\ref{fig_omangres}, for those obtained with 
$g_{\sigma N P_{11}}=-4.3$ (the upper panel) 
and $g_{\sigma N P_{11}}=+4.8$ (the lower panel).
Although the value, $g_{\sigma N P_{11}}=+4.8$ reproduces the $\omega$ 
angular distribution data from COSY-TOF~\cite{COSY} better, the result 
for the energy dependence of the total cross section is worse 
than that with $g_{\sigma NP_{11}} = -4.3$. 
Thus, we will show only the results obtained with 
$g_{\sigma N P_{11}}=-4.3$. 
We summarize in Table~\ref{tab_res} all the parameters fixed in 
the present approach, i.e., with the inclusion of the nucleon resonances.

\begin{figure}
\begin{center}
\epsfig{file=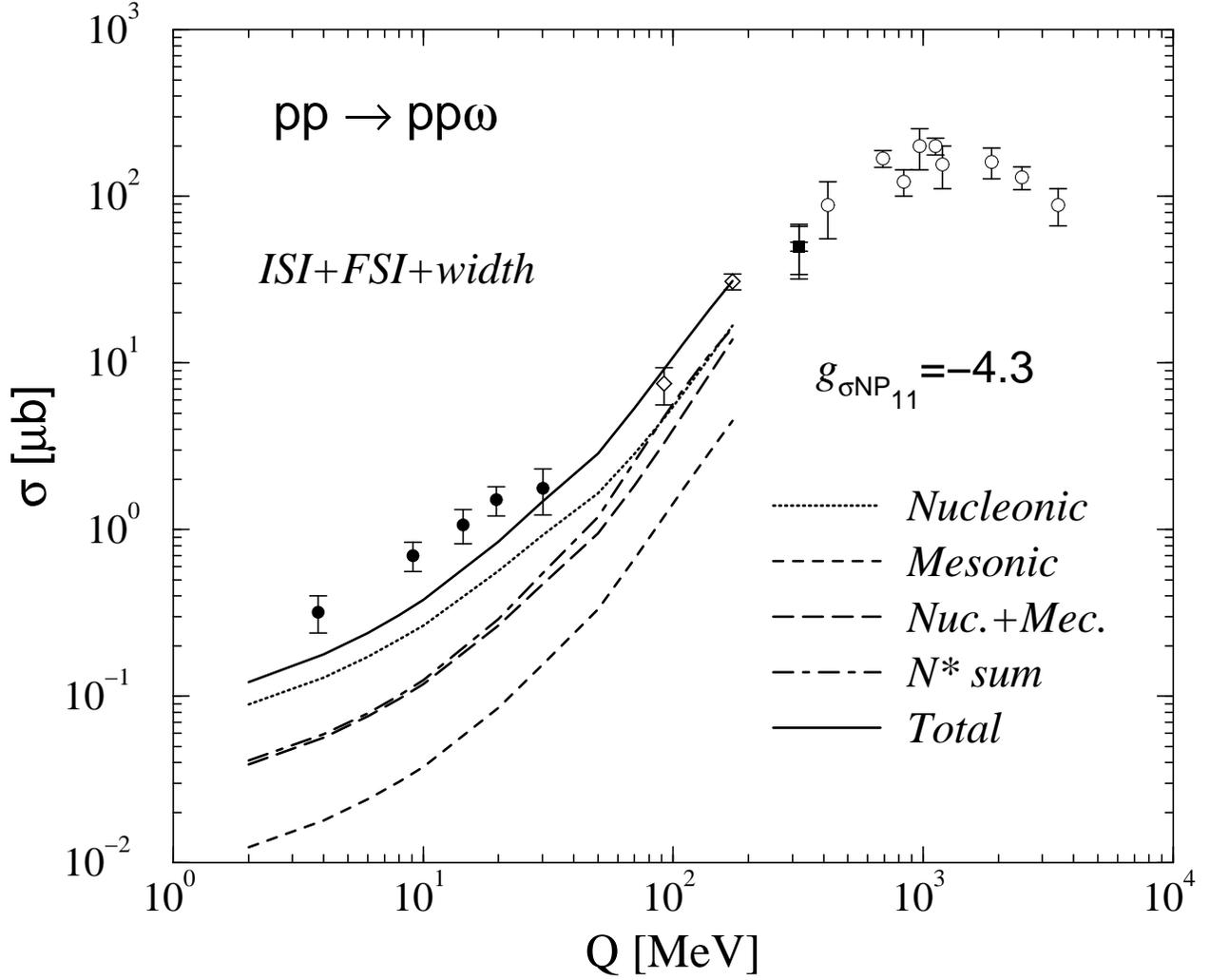,width=14cm,angle=-90}
\caption{Calculated energy dependence of the total cross section 
for the $p p \to p p\, \omega$ reaction with the inclusion of 
nucleon resonances, $S_{11}(1535), P_{11}(1710), D_{13}(1700)$ 
and $P_{13}(1720)$.
Also, see the caption of Fig.~\protect\ref{fig_noresxsec}.
}
\label{fig_xsecomega}
\end{center}
\end{figure}

Next, in Fig.~\ref{fig_xsecomega} we show the energy dependence 
of the $p p \to p p\omega$ total cross section calculated 
using the fixed parameters 
in Table~\ref{tab_res}. The result is greatly improved   
compared to that without the inclusion of any nucleon 
resonances studied in Section~III. (See Fig.~\ref{fig_noresxsec}.) 
However, it still underestimates 
the SATURNE data~\cite{Hibou}, which are in the range of 
excess energies $Q < 31$ MeV, by about a factor of two. 
Thus, further investigation is needed to 
understand better the near-threshold $p p \to p p \omega$ reaction.
As already mentioned, we also need 
more data for exclusive observables in the energy region above but close to  
$Q = 30$ MeV, because there is no established method for 
removing the multi-pion background associated with the $\omega$-meson width
from the raw data to extract the cross sections. The effect of
the width is very important in the energy region, $Q < 30$ MeV,
and the extraction of the cross section can become highly model dependent.

In order to explore the sensitivity of more exclusive observables than
cross sections to the nucleon resonances in the 
$p p \to p p \omega$ reaction, we have also calculated the spin
correlation functions and analyzing power at $Q = 92$ and $173$ MeV, 
with and without the inclusion of the nucleon resonance currents. 
Here, we just mention that although some of
the spin correlation functions exhibit some sensitivity to the presence
of nucleon resonances, judging from the currently achieved precisions 
for the $p p \to p p \omega$ data, such a presence may not be sensitive 
enough to be distinguished experimentally. A more
thorough and complete study of the role of nucleon resonances in the
production of $\omega$ mesons in $NN$ collisions, especially in a 
combined analysis of the $p p \to p p \omega$ and
$p n \to d \omega$ reactions, will be reported elsewhere~\cite{TN_inv}. 
Such an analysis will consider, not only the spin observables, 
but also the invariant mass distributions.
The $p n \to d \omega$ process, for which the total cross section 
data have been reported quite recently~\cite{COSYd},
will provide additional constraints on the model parameters.

Considering the results shown in Fig.~\ref{fig_xsecomega}, one possibility
for improving the agreement with the data would be to introduce 
extra resonances, which enhance the total cross section at near 
threshold energies but only moderately enhance at excess energies  
around $Q = 173$ MeV, if such adequate candidates exist.
On the other hand, the introduction of new resonances 
would introduce more ambiguities.
The other effect to be investigated is the 
$\omega N$ FSI, which is expected to 
enhance the total cross sections at near threshold energies, 
because a QCD sum rule study of the meson-nucleon spin-isospin averaged 
scattering lengths for the vector mesons $\rho, \omega$ and $\phi$, 
suggests attractive $VN$ ($V = \rho, \omega, \phi$) 
interactions~\cite{Koike}.   

\begin{figure}
\begin{center}
\epsfig{file=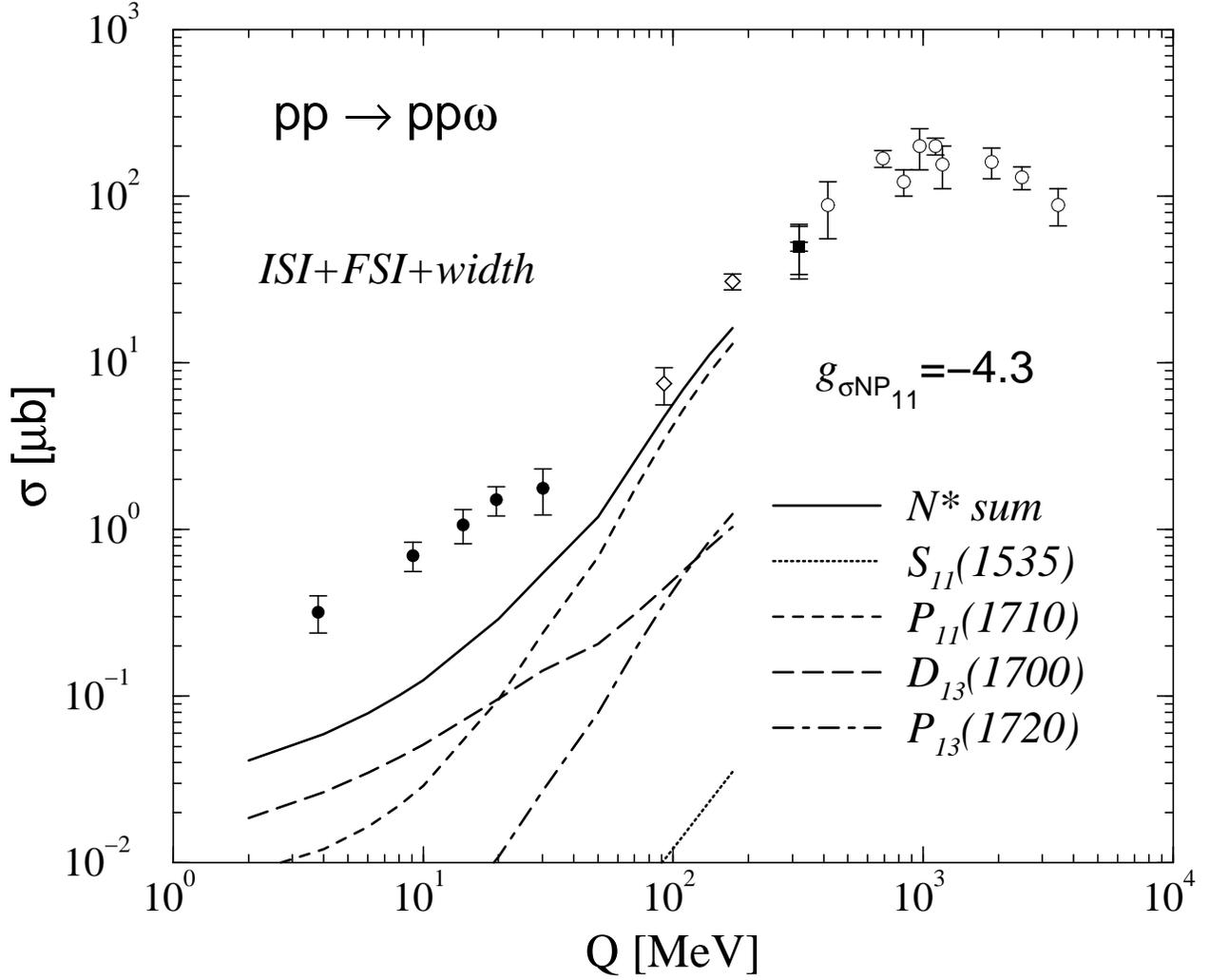,width=14cm,angle=-90}
\caption{Decomposition of resonance contributions for the energy dependence
of the $p p \to p p \omega$ total cross section. 
Also, see the caption of Fig.~\protect\ref{fig_noresxsec}.
}
\label{fig_resxsec}
\end{center}
\end{figure}

Next, in Fig.~\ref{fig_resxsec} we show a decomposition of each resonance 
contribution to the energy dependence of the 
$p p \to p p \omega$ total cross section.
Close to threshold energies the dominant contribution comes 
from $D_{13}(1700)$, while at higher excess energies, the 
dominant contribution comes from $P_{11}(1710)$, although the 
contribution of this resonance was negligible in the 
$\pi^- p \to \omega n$ reaction.
Here, again the $S_{11}(1535)$ resonance contribution 
is very small in our model.
We should mention that we also studied the contribution from 
the $S_{11}(1650)$ resonance in the present approach, 
but it did not give an appreciable contribution.  
Thus, if we want to be consistent with both the $\pi^- p \to \omega n$ and 
$p p \to p p \omega$ reactions, it appears necessary 
to include at least three nucleon resonances, 
$P_{11}(1710), D_{13}(1700)$ and $P_{13}(1720)$, 
in the present approach.

Finally, before leaving this section, we should mention that the
treatment of the $\pi^- p \to \omega n$ reaction should be improved 
in our model. In particular, as shown by
Penner and Mosel~\cite{Mosel}, effects of higher-order terms than the Born
term in the t-matrix equation is important and, consequently, 
they should be taken into account in a better way
than through a form factor as has been done in the present work.

\section{$p p \to p p \phi$ without resonance}

The $p p \to p p \phi$ reaction can be treated in an analogous way to the 
$p p \to p p \omega$ reaction in Section III, namely, considering only the  
nucleonic ($J_{nuc}^\mu$) and mesonic ($J_{mec}^\mu$) current contributions. 
However, the scarcity of data, especially in the near-threshold region,  
makes this study more difficult. In fact there is only one 
total cross section and one angular distribution
available near-threshold at an excess energy 
of $Q = 83$ MeV measured by the DISTO Collaboration~\cite{DISTO}. 
A theoretical study of the $p p \to p p \phi$ reaction in 
Ref.~\cite{Kanzo_phi} was made within a similar approach to that of 
the present study, in the sense that it used a relativistic 
meson-exchange model, considering contributions from the 
nucleonic and $\phi\rho\pi$ exchange currents as the dominant contributions. 
The present study differs from that of Ref~\cite{Kanzo_phi} in that: 
(1) the $\phi$ angular distribution data from DISTO~\cite{DISTO} 
used in this study were reanalyzed~\cite{DISTO} and absolute 
normalization of the corresponding total cross section was 
established, and (2) the $pp$ ISI is included explicitly.
 
The relative importance among the possible 
meson exchange current contributions was estimated based on an 
SU(3) effective Lagrangian, together with various 
effects described in Ref.~\cite{Kanzo_phi}. 
Furthermore, the test calculations performed in 
Ref.~\cite{Kanzo_phi} showed that the $\phi\rho\pi$-exchange current 
was by far the dominant mesonic current. 
The {\it combined} contribution of all other meson-exchange 
currents to the total cross section is about two orders of magnitude
smaller than this. Moreover, possible contributions from 
meson-exchange currents involving heavy mesons, in particular,
the $\phi\phi f_1$- and $\phi\omega f_1$-exchange currents 
were also examined using the larger values of the coupling
constants calculated from the observed decay of 
$f_1 \rightarrow \phi + \gamma$. However, this contribution, 
as well as the $\phi\phi\sigma$- and $\phi\omega\sigma$-exchange 
currents, also turned out to be negligible~\cite{Kanzo_phi}. 
Finally, as in the case of $\omega$ production, there are neither 
experimental indications of any of the known isospin-1/2 $N^*$ resonances
decaying into the $N\phi$ channel, nor do there exist enough data 
for the near-threshold $p p \to p p \phi$ reaction   
to fix the relevant parameters and judge their validities. 
Thus, we study the $p p \to p p \phi$ reaction considering 
the contributions only from the nucleonic and 
$\phi\rho\pi$ mesonic currents. 

The coupling constant, $g_{\phi\rho\pi}$, associated with the 
mesonic current $J^\mu_{mec}$, can be extracted from
the measured branching ratio. Specifically, the coupling constant 
$g_{\phi\rho\pi}=-1.64$ is determined directly from the measured 
decay width of $\phi \rightarrow \rho + \pi$~\cite{PDG}, where 
the sign is inferred from SU(3) symmetry~\cite{Kanzo_phi}. 
We note that the coupling constant $g_{\phi\rho\pi}=-1.64$ is 
extracted at different kinematics compared to that of the $\omega$: 
$g_{\phi\rho\pi}$ is determined at $q^2_\rho = m^2_\rho$ 
and $q^2_\pi = m^2_\pi$, whereas $g_{\omega\rho\pi}$ is extracted
at $q^2_\rho = 0$ and $q^2_\pi = m^2_\pi$. 
Then, the corresponding form factor (cf. Eq.~(\ref{formfactorM})) 
is defined by
\begin{equation}
F_{\phi\rho\pi}(q_\rho^2, q_\pi^2) \equiv 
F_\rho (q_\rho^2) F_\pi (q_\pi^2) = \left (\frac{\Lambda_\rho^2 - m_\rho^2}
            {\Lambda_\rho^2 - q_\rho^2}\right) 
\left (\frac{\Lambda_\pi^2 - m_\pi^2} {\Lambda_\pi^2 - q_\pi^2}\right ) \ . 
\label{formfactorM3}
\end{equation}
In Eq.~(\ref{formfactorM3}) we again assume the cut-off parameter 
$\Lambda_\phi \equiv \Lambda_\rho = \Lambda_\pi$ corresponds 
to those in Eq.~(\ref{formfactorM}).

Although we do not have to use the same parameters for 
the $\phi$ production as those used in the $p p \to p p \omega$ 
reaction, we use the same value $\Lambda_N = 1190$ MeV for 
the cut-off parameter in the $\phi NN$ meson production vertex, because 
the $\phi\rho\pi$ exchange current gives the dominant contribution 
to the total cross section, and reproducing the 
absolute normalization of the total cross section 
is relatively insensitive to the cut-off parameter $\Lambda_N$ 
in the nucleonic current compared to the cut-off parameters 
in the $\phi\rho\pi$ vertex form factors.
Therefore, we have three free parameters 
to be adjusted to reproduce the $\phi$ meson production total cross section 
and angular distribution data~\cite{DISTO}, 
namely, $g_{\phi NN}$ and $\kappa_\phi$ in the nucleonic current,  
after replacing $\omega \to \phi$ in Eq.~(\ref{NNomega}), 
and $\Lambda_\phi \equiv \Lambda_\rho = \Lambda_\pi$ 
in the form factor of Eq.~(\ref{formfactorM3}) 
in the $\phi\rho\pi$ mesonic current, after replacing $\omega \to \phi$ in 
Eqs.~(\ref{pirhoomega}) and~(\ref{mec_cur}).

\begin{figure}
\begin{center}
\epsfig{file=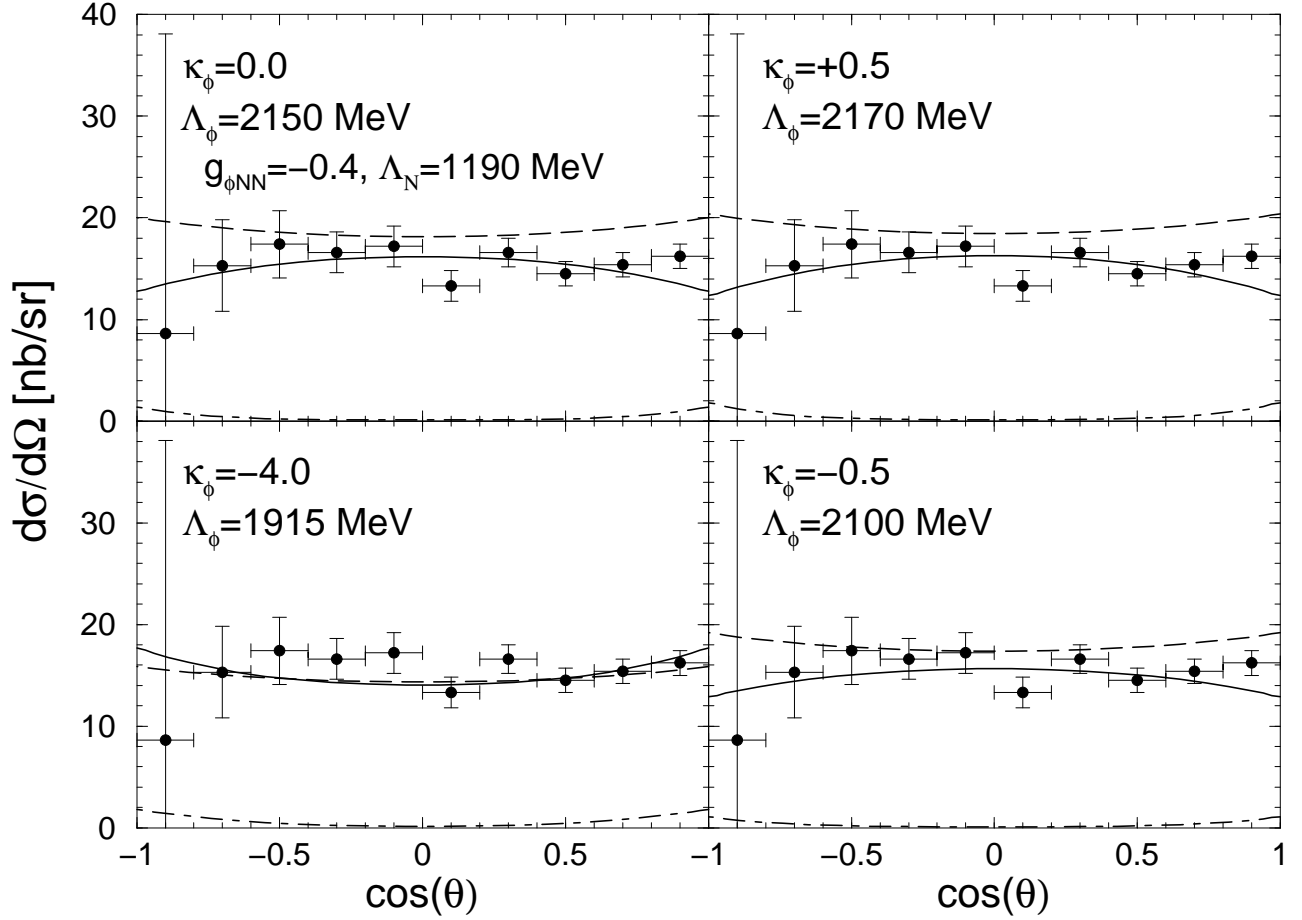,width=14cm,angle=-90}
\caption{$\kappa_\phi$ dependence of the $\phi$ angular 
distribution at $Q = 83$ MeV. The (dashed, dot-dashed, solid) lines show the  
(mesonic, nucleonic, total) contributions, respectively. 
The $\phi NN$ coupling constant and the cut off parameter 
$\Lambda_N$ associated with the $\phi NN$ meson production vertex 
are fixed at $g_{\phi NN} = -0.4$ and $\Lambda_N = 1190$ MeV, 
and the cut-off parameter $\Lambda_\phi \equiv \Lambda_\rho = \Lambda_\pi$ 
in the $\phi\rho\pi$ meson production  
vertex is fitted to reproduce the 
total cross section of $190 nb$ at $Q = 83$ MeV. 
The dots are the data from DISTO~\protect\cite{DISTO}(a). 
}
\label{fig_phiang1}
\end{center}
\end{figure}

\begin{figure}
\begin{center}
\epsfig{file=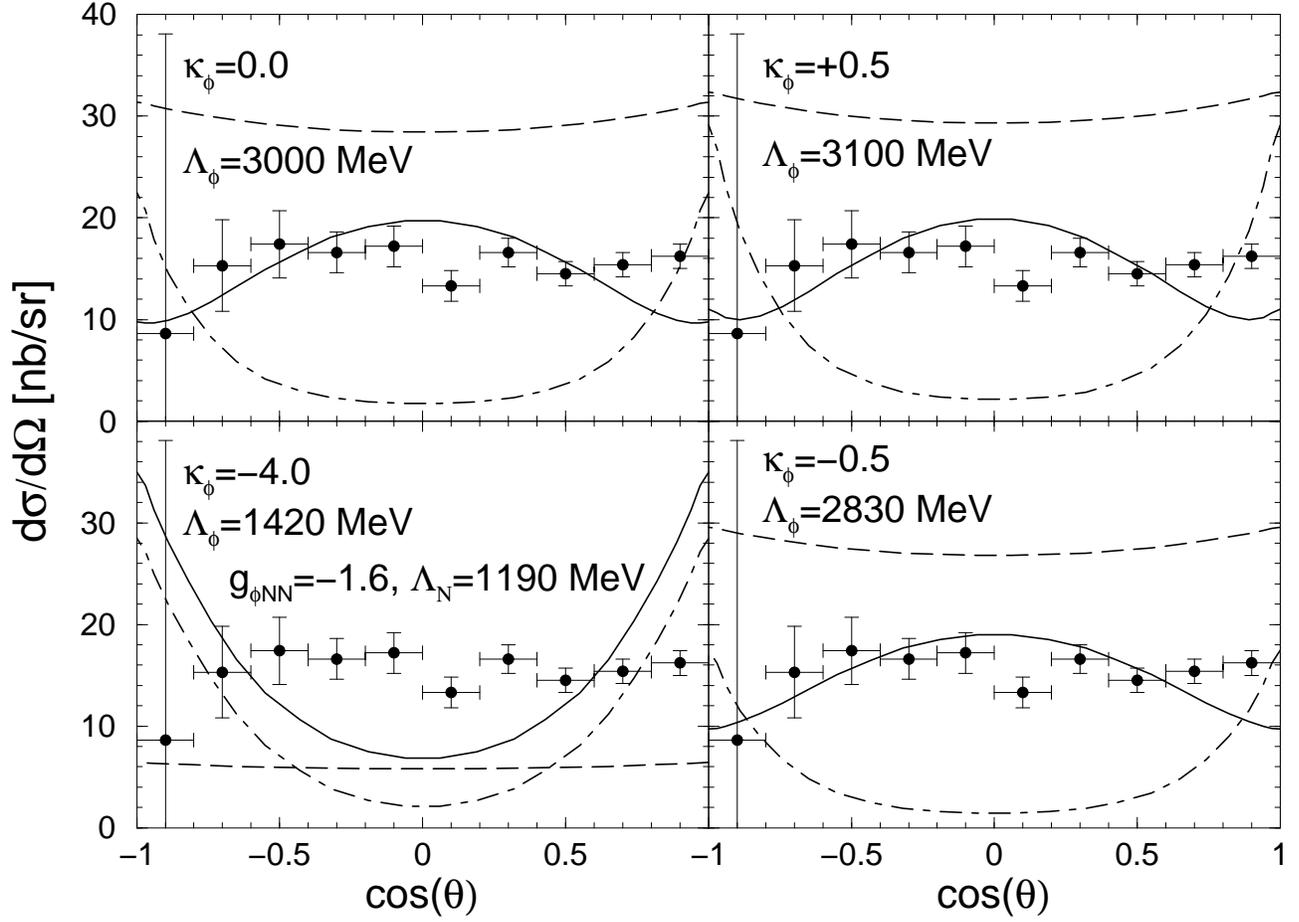,width=14cm,angle=-90}
\caption{Same as Fig.~\protect\ref{fig_phiang1}, but $g_{\phi NN} = -1.6$.
}
\label{fig_phiang2}
\end{center}
\end{figure}

In Figs.~\ref{fig_phiang1} and~\ref{fig_phiang2} we show the $\kappa_\phi$ 
dependence of the calculated $\phi$ angular distributions, for 
$g_{\phi NN}=-0.4$ and $g_{\phi NN}=-1.6$, respectively.
Results in Fig.~\ref{fig_phiang1} imply that as long as the value of 
$g_{\phi NN}$ is small, the angular distribution data can be reproduced 
well within experimental error bars, irrespective of the values of 
$\kappa_\phi$ up to $\kappa_\phi \simeq -4.0$. 
On the other hand, results in Fig.~\ref{fig_phiang2} 
show that the larger value, $g_{\phi NN} = -1.6$ makes   
the shape of the calculated $\phi$ angular distribution sensitive 
to $\kappa_\phi$. 
One can notice that for a certain value of $\kappa_\phi$, the shape of   
the calculated $\phi$ angular distribution changes from convex 
to concave.
After some test calculations, we find the optimum value for this transition  
is roughly $\kappa_\phi \simeq -2.0$. Thus, around 
$\kappa_\phi \simeq -2.0$ we can expect that there are a large number 
of possibilities for the values of $g_{\phi NN}$ and $\Lambda_\phi$ 
which can reproduce the experimentally observed flat 
$\phi$ angular distribution~\cite{DISTO}.

\begin{figure}
\begin{center}
\epsfig{file=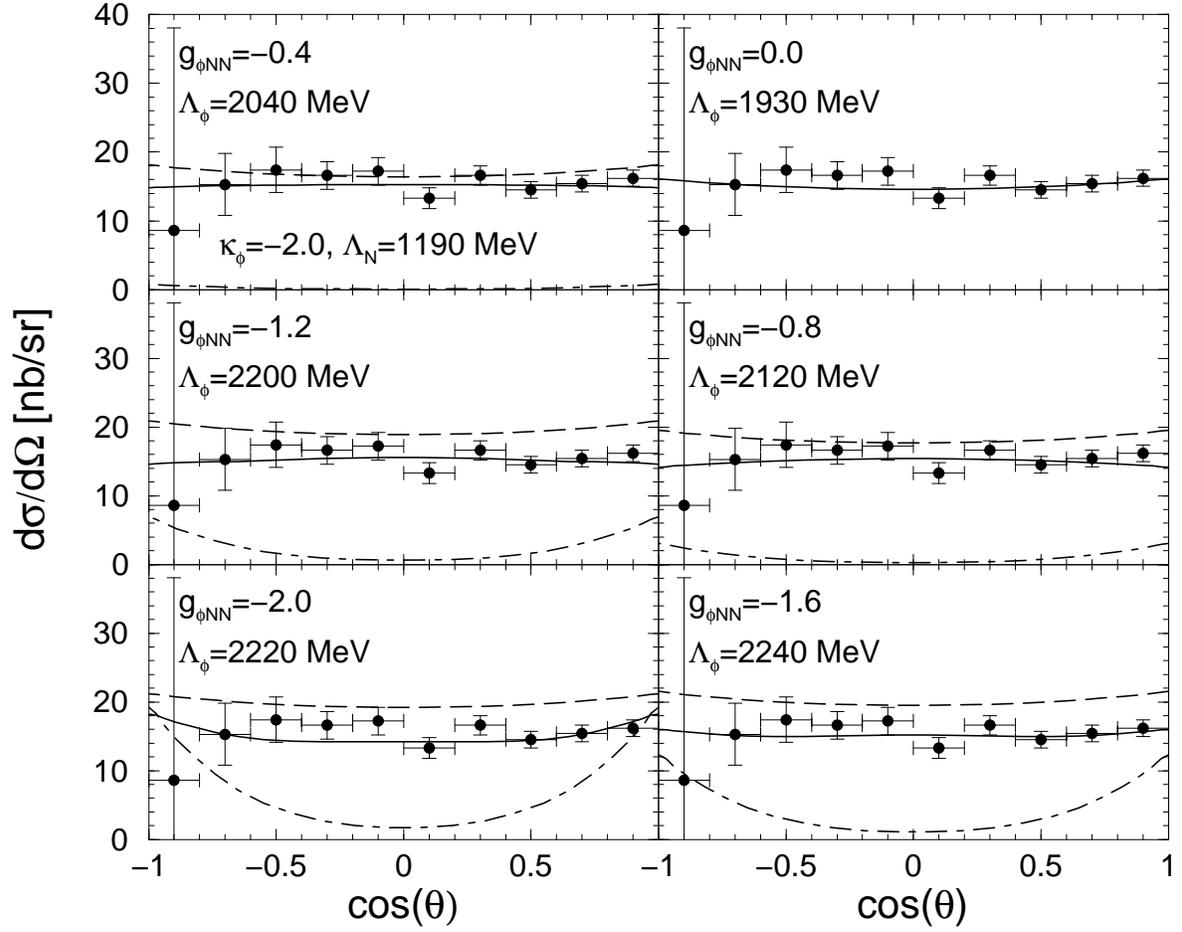,width=14cm,angle=-90}
\caption{$g_{\phi NN}$ dependence of the $\phi$ angular 
distribution with the fixed value $\kappa_\phi = -2.0$, and 
$\Lambda_N = 1190$ MeV.
Also, see the caption of Fig.~\protect\ref{fig_phiang1}.
}
\label{fig_phiang3}
\end{center}
\end{figure}

Next, we fix the value $\kappa_\phi = -2.0$ (and $\Lambda_N = 1190$ MeV), 
and study the $g_{\phi NN}$ dependence of the $\phi$ angular distribution.  
Some of the calculated results are shown in Fig.~\ref{fig_phiang3}. 
Note that, the top-right panel in Fig.~\ref{fig_phiang3} has a 
contribution solely from the mesonic current ($g_{\phi NN}=0$), 
which, neglecting the $\phi - \omega$ mixing and the OZI-allowed 
two step processes~\cite{Kanzo_phi}, 
may be regarded as the limiting case of no $s \bar{s}$ 
component in the nucleon wave function,
if the value of $g_{\phi NN}$ is considered as a measure 
for the $s \bar{s}$ component.
Since we have not included quark degrees of freedom explicitly, 
it is difficult to draw a definite conclusion on the $s \bar{s}$ 
component in the nucleon wave function. 
We summarize in Table~\ref{tab_phi} the four possible parameter sets for
$g_{\phi NN}, \kappa_\phi$ and $\Lambda_\phi$ fixed by
fitting the DISTO $\phi$ angular distribution data~\cite{DISTO}.
They all reproduce the experimental data~\cite{DISTO} reasonably well.
This suggests that one needs to study additional observables, 
e.g., the energy dependence of the $p p \to p p \phi$ total cross section, 
in order to constrain better the parameters of the model. 

Here, it may be interesting to compare the values of $g_{\phi NN}$ and 
$\kappa_\phi$ obtained in the present work with those 
extracted in Ref.~\cite{Cotanch} by studying  
the off-shell time-like nucleon form factors using 
the $p(\gamma ,e^+e^-)p$ reaction. They obtained  
$(g_{\phi NN}, \kappa_\phi) \simeq (1.3, 7.2)$.
(Note that their definition of $\kappa_\phi$ is different from 
that of the present study by a factor $4m_N/m_\phi$ 
($m_{N,\phi}$: masses of the nucleon and $\phi$ meson); for comparison, 
this factor is included in the value of $\kappa_\phi$ here.)

\begin{figure}
\begin{center}
\epsfig{file=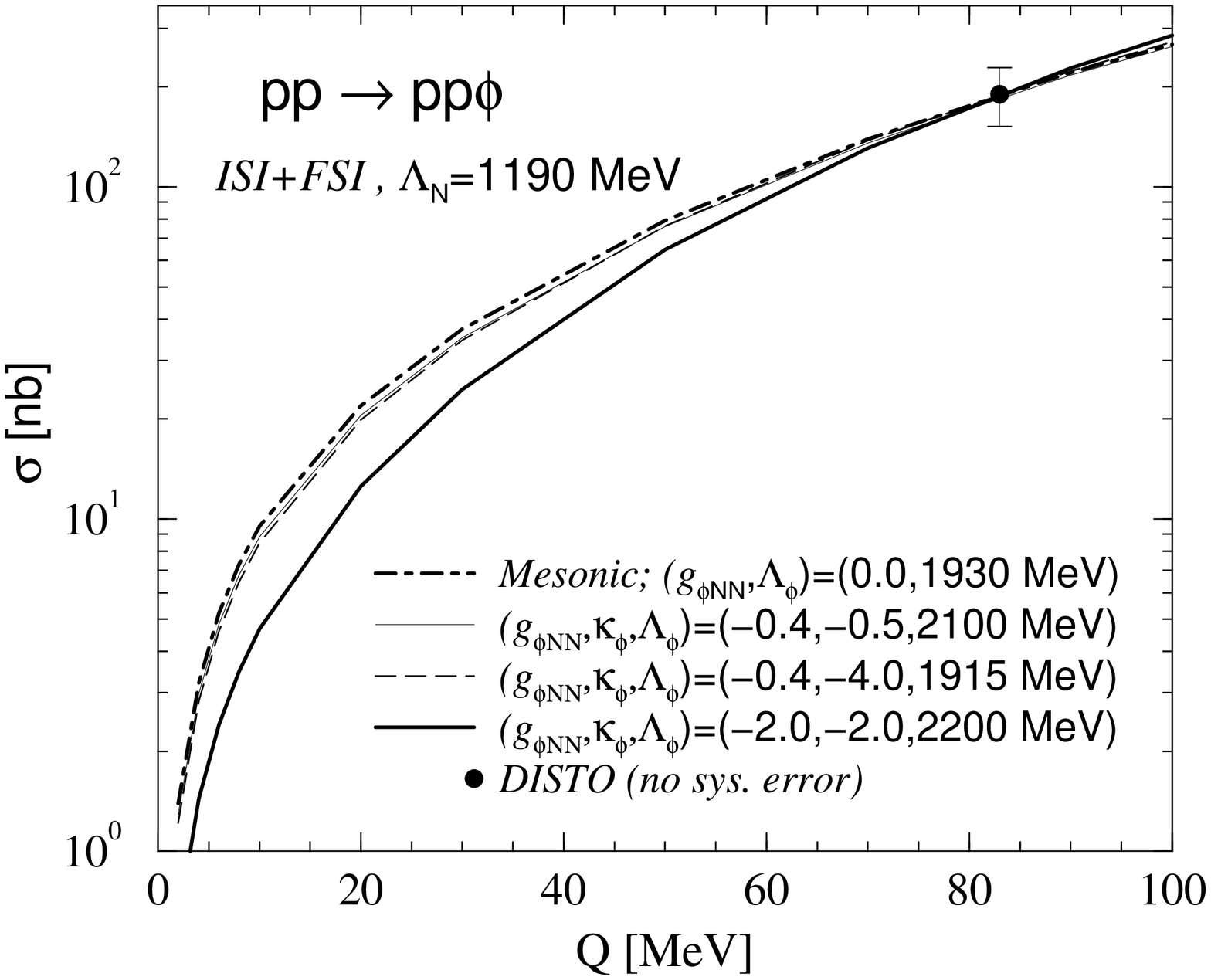,width=14cm,angle=-90}
\caption{Energy dependence of the total cross section for the 
$ p p \to p p \phi$ reaction, calculated using the four 
parameter sets, which can reproduce the $\phi$ angular distribution 
data~\protect\cite{DISTO}.
A systematic error in the data point from DISTO~\protect\cite{DISTO} 
is not included.
}
\label{fig_phixsec}
\end{center}
\end{figure}

Next, using the four parameter sets given in Table~\ref{tab_phi}, we study  
the energy dependence of the $p p \to p p \phi$ total cross section.
We show the calculated results in Fig.~\ref{fig_phixsec}.
The results exhibit very similar energy dependences for the parameter sets
$(g_{\phi NN}, \kappa_\phi) = (0.0, 0.0), (-0.4, -0.5)$ and 
$(-0.4, -4.0)$, while that for $(-2.0, -2.0)$ 
shows a different dependence especially at excess energies in the region 
$Q < 50$ MeV. 
Thus, measuring the energy dependence of the total cross section for 
$Q < 50$ MeV will help constrain better the model parameters, 
in particular, the magnitude of the coupling constant  
$g_{\phi NN}$. 

\section{Summary and discussion}

We have studied the $p p \to p p \omega/\phi$ reactions  
using a relativistic effective Lagrangian at the hadronic level, 
including both the initial and final state $pp$ interactions.
For both reactions we have made use of the recently measured 
$\omega$ and $\phi$ angular distributions 
in addition to the total cross section data to fix the model parameters. 

We have studied the $p p \to p p \omega$ reaction considering 
two possibilities, i.e., the $\omega$ meson is produced by:  
(1) the nucleonic and mesonic current contributions, and 
(2) the nucleonic, mesonic and nucleon resonance current contributions.
The results show that the energy dependence of the total cross section 
in the range of excess energies $Q \le 173$ MeV, is apparently described 
better by the inclusion of nucleon resonances, 
which is implemented in a way to be consistent with the 
$\pi^- p \to \omega n$ reaction. However, the calculation still underestimates 
the SATURNE data by about a factor of two, where the data points are 
in the range of excess energies $Q < 31$ MeV. 
This remains still a problem in understanding the reaction mechanism. 
In this connection, we need more data for 
exclusive observables in the energy region above, but close to
$Q = 30$ MeV because there is no established method for 
removing the multi-pion background associated with the $\omega$-meson width
from the raw data. This removal is necessary for extracting 
the cross sections in the energy region, $Q < 30$ MeV, where the effect of 
the width is very important, and the extraction can be highly model dependent.


In connection with studying resonance contributions to the 
$p p \to p p \omega$ reaction, we plan to investigate  
the $p\omega$ invariant mass distributions for this reaction~\cite{TN_inv}.
A measurement of the invariant mass distributions 
for this reaction, should give  
significant information as to whether contributions from resonances  
are appreciable or not. Such theoretical studies have been made for  
the $p p \to p p \eta$~\cite{Kanzo_inv}, and  
$p p \to p \Lambda K^+$~\cite{AlexK} reactions.
Thus, the study of the invariant mass distributions 
may be an alternative method for studying   
the possible $\omega$-meson (and $\phi$-meson) 
resonance couplings both theoretically and experimentally. 

In addition to the $p p \to p p \omega$ reaction, 
we have studied the $p p \to p p \phi$ reaction 
considering the contributions solely from 
the nucleonic and mesonic current contributions. 
Because of the scarcity of data for this reaction     
in the near threshold energy region, 
we have obtained four parameter sets which can  
reproduce the $\phi$ angular distribution data from 
DISTO~\cite{DISTO} equally well.
Predictions for the energy dependence of the $p p \to p p \phi$ 
total cross section indicate that a measurement of 
the cross section close to threshold should be able to 
constrain better the coupling constant $g_{\phi NN}$ 
($\simeq 0$ or $\simeq -2$). 

Finally, although there exists an enormous interest in 
vector meson properties in highly complicated many-nucleon environments, 
e.g., dilepton production in heavy ion 
collisions and meson ($\omega$) nuclear bound states,  
the data for the $N N \to N N V$ reaction 
($V$: vector meson) are currently inadequate 
for understanding the production 
mechanism of these mesons in free space. 
Thus, more measurements of vector meson 
production in free space, and especially in $NN$ collisions,  
may be a first step towards understanding the properties of vector 
mesons in such complicated nuclear environments.
  
\vspace{2em}
\noindent
{\bf Acknowledgment:}\\
\noindent
We would like to thank C. Wilkin and F. Hibou for useful 
discussions, and providing us the code used in the analysis of 
the SATURNE data~\cite{Hibou}. Our thanks also go to K.-Th. Brinkmann 
for providing us the $\omega$ angular distribution data of 
the COSY-TOF Collaboration~\cite{COSY},  
and to J. Haidenbauer for helpful discussions.
We also thank W.G. Love for a careful reading of the manuscript.
This work is supported by Forschungszentrum-J\"{u}lich, contract No. 
41445282 (COSY-058).



%
\newpage
\begin{table}[hbtp]
\begin{center}
\caption{Model parameters fixed for the $p p \to p p \omega$ reaction
without the inclusion of nucleon resonances.
Below, ``Bonn" indicates that the same value  
in the Bonn $NN$ potential B (Table A.1)~\protect\cite{Bonn} is used.}
\begin{tabular}{lll}
Vertex &Coupling constant &Cut-off (MeV)\\
\hline\hline
Nucleonic current: &$[f_{\omega NN} = \kappa_\omega g_{\omega NN}]$ & \\
$\omega NN$ [$\omega$ production] &$g_{\omega NN}=9.0$ 
&$\Lambda_N=1190$ [See Eq.~(\ref{formfactorN}).]\\
                       &[$\kappa_\omega=-2.0$] & \\
$MNN [M=\pi,\eta,\rho,\omega,\sigma,a_0(=\delta)]$ &Bonn &Bonn\\
\hline\hline
Mesonic current: & & \\
$\omega\rho\pi$ [$\omega$ production] &$g_{\omega\rho\pi}=10.0$ 
&$\Lambda_\omega \equiv \Lambda_\rho = \Lambda_\pi=1000$ 
[See Eq.~(\ref{formfactorM}).]\\
$\rho NN$ &Bonn &Bonn\\
$\pi NN$ [pv-coupling] &Bonn &1300\\
\end{tabular}
\label{tab_nores}
\end{center}
\end{table}
\newpage
\begin{table}[hbtp]
\begin{center}
\caption{Model parameters fixed for the $p p \to p p \omega$ reaction 
with the inclusion of nucleon resonances.
Below, ``Bonn" indicates that the same value           
in the Bonn $NN$ potential B (Table A.1)~\protect\cite{Bonn} is used.}
\begin{tabular}{lll}
Vertex &Coupling constant &Cut-off (MeV)\\
\hline\hline
Nucleonic current: &$[f_{\omega NN} = \kappa_\omega g_{\omega NN}]$ & \\
$\omega NN$ [$\omega$ production] &$g_{\omega NN}=9.0$
&$\Lambda_N=1100$ [See Eq.~(\ref{formfactorN}).]\\
                       &[$\kappa_\omega=-0.5$] & \\
$MNN [M=\pi,\eta,\rho,\omega,\sigma,a_0(=\delta)]$ &Bonn &Bonn\\
\hline\hline
Mesonic current: & & \\
$\omega\rho\pi$ [$\omega$ production] &$g_{\omega\rho\pi}=10.0$
&$\Lambda_\rho=850, \Lambda_\pi=1450$ [See Eq.~(\ref{formfactorM2}).]\\
$\rho NN$ &Bonn &Bonn\\
$\pi NN$ [pv-coupling] &Bonn &900\\
\hline\hline
Spin 1/2 resonance current: &$g_{MNN},g_{MNN^*}$ & \\
 &$\frac{1}{m_{N^*}+m_N}(g_{\rho, \omega NN^*},\, 
\kappa_{\rho, \omega} g_{\rho, \omega NN^*})$ 
&[See Eqs.~(\ref{NR12omega}) and~(\ref{NR12rho}).]\\ 
$S_{11}(1535),\Gamma=150$ MeV & & \\
$MNN [M=\pi,\eta,\rho,\omega]$ &Bonn, but $g_{\omega NN} = 9.0$ 
&Bonn, but $\Lambda_{\pi NN} = 900$\\
$[\pi NN$ (pv-coupling)] & & \\
$\pi NS_{11}(1535)$    &1.25 &900\\
$\eta NS_{11}(1535)$   &2.02 &Bonn\\
$\rho NS_{11}(1535)$   &(0.0, -4.50)  [fm] &Bonn\\
$\omega NS_{11}(1535)$ &(-1.04, 3.82) [fm] &Bonn\\
\hline
$P_{11}(1710),\Gamma=100$ MeV & & \\
$MNN [M=\sigma,\pi,\eta,\rho,\omega]$ &Bonn, but $g_{\omega NN} = 9.0$ 
&Bonn, but $\Lambda_{\pi NN} = 900$\\
$[\pi NN$ (pv-coupling)] & & \\
$\sigma NP_{11}(1710)$ &-4.30&Bonn\\
$\pi NP_{11}(1710)$    &1.20 &900\\
$\eta NP_{11}(1710)$   &4.43 &Bonn\\
$\rho NP_{11}(1710)$   &(0.0, 6.70)  [fm] &Bonn\\
$\omega NP_{11}(1710)$ &(0.0, -1.19) [fm] &Bonn\\
\hline\hline
Spin 3/2 resonance current: &$g_{MNN},g^{(1)}_{MNN^*}$ & \\
 &[$g^{(1)}_{\rho ,\omega NN^*}/g^{(2)}_{\rho ,\omega NN^*}=-2.1$]   
 &[See Eqs.~(\ref{NR32omega}) and~(\ref{NR32rho}).] \\
$D_{13}(1700),\Gamma=100$ MeV & &\\
$MNN [M=\pi,\rho,\omega]$ &Bonn, but $g_{\omega NN} = 9.0$
&Bonn, but $\Lambda_{\pi NN} = 900$\\
$[\pi NN$ (pv-coupling)] & & \\
$\pi ND_{13}(1700)$    &0.44 &900\\
$\rho ND_{13}(1700)$   &1.68 &Bonn\\
$\omega ND_{13}(1700)$ &3.02 &Bonn\\
\hline
$P_{13}(1720),\Gamma=150$ MeV & &\\
$MNN [M=\pi,\rho,\omega]$ &Bonn, but $g_{\omega NN} = 9.0$
&Bonn, but $\Lambda_{\pi NN} = 900$\\
$[\pi NN$ (pv-coupling)] & & \\
$\pi NP_{13}(1720)$    &0.17 &900\\
$\rho NP_{13}(1720)$   &-3.73&Bonn\\
$\omega NP_{13}(1720)$ &3.94 &Bonn\\
\end{tabular}
\label{tab_res}
\end{center}
\end{table}
\newpage
\begin{table}[hbtp]
\begin{center}
\caption{Model parameters for the $p p \to p p \phi$ reaction,  
for four possible sets by the $\phi$ angular distribution, 
denoted by (a), (b), (c) and (d). 
Below, ``Bonn" indicates that the same value           
in the Bonn $NN$ potential B (Table A.1)~\protect\cite{Bonn} is used.}
\begin{tabular}{lll}
Vertex &Coupling constant &Cut-off (MeV)\\
\hline\hline
Nucleonic current: &$[f_{\phi NN} = \kappa_\phi g_{\phi NN}]$ & \\
$\phi NN$ [$\phi$ production] &(a) $g_{\phi NN}=0.0\, [\kappa_\phi \equiv 0]$
&$\Lambda_N=1190$ [See Eq.~(\ref{formfactorN}).]\\
  &(b) $g_{\phi NN}=-0.4\, [\kappa_\phi=-0.5]$ &$\Lambda_N=1190$\\
  &(c) $g_{\phi NN}=-0.4\, [\kappa_\phi=-4.0]$ &$\Lambda_N=1190$\\
  &(d) $g_{\phi NN}=-2.0\, [\kappa_\phi=-2.0]$ &$\Lambda_N=1190$\\
$MNN [M=\pi,\eta,\rho,\omega,\sigma,a_0(=\delta)]$ &Bonn &Bonn\\
\hline\hline
Mesonic current: & &$\Lambda_\phi \equiv \Lambda_\rho = \Lambda_\pi$ \\
(a) $\phi\rho\pi$ [$\phi$ production] &$g_{\phi\rho\pi}=-1.64$
&$\Lambda_\phi=1930$ [See Eq.~(\ref{formfactorM3}).]\\
(b) $\phi\rho\pi$ [$\phi$ production] &$g_{\phi\rho\pi}=-1.64$
&$\Lambda_\phi=2100$\\
(c) $\phi\rho\pi$ [$\phi$ production] &$g_{\phi\rho\pi}=-1.64$
&$\Lambda_\phi=1915$\\
(d) $\phi\rho\pi$ [$\phi$ production] &$g_{\phi\rho\pi}=-1.64$
&$\Lambda_\phi=2200$\\
$\rho NN$ &Bonn &Bonn\\
$\pi NN$ [pv-coupling] &Bonn &1300\\
\end{tabular}
\label{tab_phi}
\end{center}
\end{table}
%
\newpage
\newpage
\newpage
\newpage
\newpage
\newpage
\newpage
\newpage
\newpage
\newpage
\newpage
\newpage
\newpage

\begin{thebibliography}{99}
%
\bibitem{Moskal}
For a recent review on close-to-threshold meson production in hadronic
interactions, see, e.g.,
P. Moskal, M. Wolke, A. Khoukaz, W. Oelert,
Part. Nucl. Phys. {\bf 49}, 1 (2002);\\
H. Machner and J. Haidenbauer, J. Phys. {\bf G 25}, R231 (1999).
%
\bibitem{Kanzo_Cracow}
K. Nakayama, nucl-th/0108032, in the Proceedings of the Symposium on
``Threshold Meson Production in $pp$ and $pd$ Interaction",
Schriften des Forschungszentrums J\"{u}lich, Matter and Materials,
{\bf 11}, 119 (2002).
%
\bibitem{Leepion}
T.-S.H. Lee and D.O. Riska, Phys. Rev. Lett. {\bf 70}, 2237 (1993);\\
C.J. Horowitz, H.O. Meyer, and D.K. Griegel, 
Phys. Rev. C {\bf 49}, 1337 (1994);\\
C. Hanhart, J. Haidenbauer, and J. Speth, 
Nucl. Phys. {\bf A631}, 515c (1998); 
C. Hanhart et al., Phys. Lett. B {\bf 358}, 21 (1995);\\
Y. Maeda, N. Matsuoka, and K. Tamura, Nucl. Phys. {\bf A684}, 392c (2001).
%
\bibitem{OZI}
S. Okubo, Phys. Lett. {\bf B5}, 165 (1963);\\
G. Zweig, CERN Report No.8419/TH412 (1964);\\
I. Iizuka, Prog. Theor. Phys. Suppl. {\bf 37 \& 38}, 21 (1966). 
%
\bibitem{OZI_review}
For a review, e.g., V.P. Nomokonov, M.G. Sapozhnikov, 
hep-ph/0204259.
%
\bibitem{Lipkin}
H.J. Lipkin, Phys. Lett. {\bf 60B}, 371 (1976).
%
\bibitem{Ellis}
J. Ellis, M. Karliner, D.E. Kharzeev, M.G. Sapozhnikov,
Phys. Lett. B {\bf 353}, 319 (1995); Nucl. Phys. {\bf A673}, 256 (2000).
%
\bibitem{ppbar}
C. Amster, Rev. Mod. Phys. {\bf 70}, 1293 (1998).
%
\bibitem{DISTO}
(a) DISTO Collaboration, F. Balestra et al., 
Phys. ReV. C {\bf 63}, 024004 (2001); 
(b) Phys. Rev. Lett. {\bf 81}, 4572 (1998).
%
\bibitem{Isgur}
N. Isgur, G. Karl, Phys. Rev. D {\bf 18}, 4187 (1978);
{\it ibid.} {\bf 19}, 2653 (1979); {\it ibid.} {\bf 23}, 817(E) (1981);\\
R. Koniuk, N. Isgur, {\it ibid.} {\bf 21}, 1868 (1980).
%
\bibitem{Capstick}
S. Capstick, W. Roberts, Phys. Rev. D {\bf 49}, 4570 (1994); 
nucl-th/0008028, submitted to Prog. Part. Nucl. Phys.;\\
S. Capstick, Phys. Rev. D {\bf 46}, 2864 (1992).
%
\bibitem{Zhao}
Q. Zhao, Z. Li, C. Bennhold, Phys. Lett. B {\bf 436}, 42 (1998);
Phys. Rev. C {\bf 58}, 2393 (1998).
%
\bibitem{Oh}
Y. Oh, A.I. Titov, T.-S.H. Lee, Phys. Rev. C {\bf 63}, 025201 (2001).
%
\bibitem{Riska}
D.O. Riska, G.E. Brown, Nucl. Phys. {\bf A679}, 577 (2001).
%
\bibitem{Mosel}
G. Penner, U. Mosel, Phys. Rev. C {\bf 65}, 055202 (2002);
nucl-th/0207066.
%
\bibitem{Titov}
A.I. Titov, T.-S.H. Lee, Phys. Rev. C {\bf 66}, 015204 (2002).
%
\bibitem{dilepton}
G. Agakichiev et al. (CERES Coll.), Phys. Rev. Lett. {\bf 75}, 1272 (1995); 
Phys. Lett. B{\bf 422}, 405 (1998);\\ 
B. Lenkeit et al. (CERES Coll.), Nucl. Phys. {\bf A661} 23c (1999);\\
M. Masera (HELIOS-3 Coll.), Nucl. Phys. {\bf A590}, 93c (1995);\\
W.K. Wilson et al. (The DLS Coll.), Phys. Rev. C {\bf 57}, 1865 (1998).
%
\bibitem{Brown}
G.E. Brown, M. Rho, Phys. Rev. Lett. {\bf 66}, 2720 (1991).
%
\bibitem{Hatsuda}
T. Hatsuda, S.H. Lee, Phys. Rev. C {\bf 46}, R34 (1992);
T. Hatsuda, S.H. Lee, and H. Shiomi, Phys. Rev. C {\bf 52}, 3364 (1995).
%
\bibitem{Asakawa}
M. Asakawa, C.M. Ko, P. L$\acute{e}$vai, X.J. Qiu, 
Phys. Rev. C {\bf 46}, R1159 (1992);\\
M. Asakawa and C.M. Ko, Phys. Rev. C {\bf 48}, R526 (1993). 
%
\bibitem{Koike}
Y. Koike and A. Hayashigaki, Prog. Theor. Phys. {\bf 98}, 631 (1997).
%
\bibitem{Klingl1}
F. Klingl, T. Waas, W. Weise, Phys. Lett. B {\bf 431}, 254 (1998).
%
\bibitem{Tsushima}
K. Tsushima, D.H. Lu, A.W. Thomas, K. Saito, 
Phys. Lett. B {\bf 443}, 26 (1998);\\
K. Saito, K. Tsushima, D.H. Lu, and A.W. Thomas,
Phys. Rev. C {\bf 59}, 1203 (1999);\\
K. Tsushima, Nucl. Phys. {\bf A670}, 198c (2000);
in the proceedings ISHEPP 98, Dubna, Russia, 17 - 22 
August 1998, nucl-th/9811063;
K. Tsushima et al., Nucl. Phys. {\bf A680}, 279 (2001);
K. Tsushima, hep-ph/0206069,  
Proceedings of the Joint CSSM/JHF/NITP Workshop on Physics at JHF,
Adelaide, Australia, 2002, p. 303.
%
\bibitem{Hayano}
R.S. Hayano, S. Hirenzaki, A. Gillitzer,
Eur. Phys. J A {\bf 6}, 99 (1999).
%
\bibitem{Klingl2}
F. Klingl, T. Waas, W. Weise, Nucl. Phys. {\bf A650}, 299 (1999).
%
%
\bibitem{Alex}
A.A. Sibirtsev, Nucl. Phys. {\bf A604}, 455 (1996).
%
\bibitem{Kanzo_omega}
K. Nakayama et al., Phys. Rev. C {\bf 57}, 1580 (1998).
%
\bibitem{Kaiser}
N. Kaiser, Phys. Rev. C {\bf 60}, 057001 (1999).
%
\bibitem{Fuchs}
C. Fuchs et al., Phys. Rev. C {\bf 67}, 025202 (2003).
%
\bibitem{TN}
K. Tsushima, and K. Nakayama, nucl-th/0211065, to be published in the 
proceedings of the XVIth International Conference on Particles and Nuclei
(PANIC02), Osaka, Japan, 30 September - 4 October 2002.
%
\bibitem{Kanzo_phi}
K. Nakayama et al., Phys. Rev. C {\bf 60}, 055209 (1999).
%
\bibitem{Titov2}
A.I. Titov, B. K\"{a}mpfer, and V.V. Shklyar, 
Phys. Rev. C {\bf 59}, 999 (1999);\\
A.I. Titov, B. K\"{a}mpfer and B.L. Reznik, 
Eur. Phys. J. {\bf A7}, 543 (2000).
%
\bibitem{Kanzo_d}
K. Nakayama, J. Haidenbauer, and J. Speth,
Phys. Rev. C {\bf 63}, 015201 (2000); Nucl. Phys. {\bf A689}, 402c (2001).
%
\bibitem{Grishina} 
V. Yu. Grishina, L.A. Kondratyuk, and M. B\"{u}scher,
Phys. Atomic Nuclei, {\bf 63}, 1824 (2000).
%
\bibitem{Wurzinger}
R. Wurzinger et al., Phys. Rev. C {\bf 51}, R443 (1995).
%
\bibitem{Flamino}
V. Flamino et al., CERN preprint CERN-HERA 8401 (1984);\\
R. Baldi et al., Phys. Lett. {\bf 68B }, 381 (1977).
%
%
\bibitem{Hibou}
F. Hibou et al., Phys. Rev. Lett. {\bf 83}, 492 (1999);\\
F. Hibou and C. Willkin, private communications.
%
\bibitem{COSY}
S. Abd El-Samad et al. (COSY-TOF Coll.),
Phys. Lett. B {\bf 522}, 16 (2001).
%
\bibitem{Kanzo_eta}
K. Nakayama, J. Speth and T.-S. H. Lee, 
Phys. Rev. C {\bf 65}, 045210 (2002).
%
\bibitem{Hanhart}
C. Hanhart and K. Nakayama, Phys. Lett. B {\bf 454}, 176 (1999).
%
\bibitem{Bonn}
J. Haidenbauer, private communication; We use the modified version of 
the Bonn-B $pp$ potential, which is constrained to the low-energy $pp$ 
scattering length by a minor readjustment of the coupling constant 
$g_{\sigma NN}$, following Phys. Rev. C {\bf 40}, 2465 (1989);\\
R. Machleidt, potential B (Table A.1), Adv. Nucl. Phys. {\bf 19}, 189 (1989). 
%
\bibitem{Blank}
R. Blankenbecler and R. Sugar, Phys. Rev. {\bf 142}, 1051 (1966). 
%
\bibitem{CNS}
R.A. Arndt, W.J. Briscoe, R.L. Workman, and I.I. Strakowsky, 
The CNS Data Analysis Center, http://gwdac.phys.gwu.edu.
%
\bibitem{Lee}
M. Batini$\acute{c}$, A. $\check{S}$varc, and T.-S.H. Lee, 
Phys. Acripta {\bf 56}, 321 (1997).
%
\bibitem{Baru}
V. Baru et al., Phys. Rev. C {\bf 67}, 024002 (2003). 
%
\bibitem{Kanzo_etap}
K. Nakayama et al., Phys. Rev. C {\bf 61}, 024001 (1999).
%
\bibitem{Pearce} 
B. C. Pearce and I. R. Afnan,
Phys. Rev. C {\bf 34}, 991 (1986).
%
%
\bibitem{Durso}
J.W. Durso, Phys. Lett. B {\bf 184}, 348 (1987). 
%
\bibitem{Garcilazo}
H. Garcilazo and E. Moya de Guerra, Nucl. Phys. {\bf A562}, 521 (1993). 
%
\bibitem{Ulf}
V. Bernard, N. Kaiser, and Ulf-G. Meissner, Int. J. Mod. Phys.
{\bf E4}, 193 (1995).
%
\bibitem{Kondratyuk}
For example, S. Kondratyuk and O. Scholten, 
Phys. Rev. C {\bf 59}, 1070 (1999).
%
\bibitem{JJHF}
K. Tsushima, hep-ph/0206069, Section 6, 
Proceedings of the Joint CSSM/JHF/NITP Workshop on Physics at JHF,
Adelaide, Australia, 2002, p. 303.
%
\bibitem{piNdata}
J.S. Danburg et al., Phys. Rev. D {\bf 2}, 2564 (1970);\\
D.M. Binnie et al., Phys. Rev. D {\bf 8}, 2789 (1973);\\
J. Keyne et al., Phys. Rev. D {\bf 14}, 28 (1976);\\
H. Karami et al., Nucl. Phys. {\bf B154}, 503 (1979).
%
\bibitem{Jlab}
M.K. Jones et al. (The Jeffereson Lab Hall A Coll.), 
Phys. Rev. Lett. {\bf 84}, 1398 (2000);\\
O. Gayou et al. (The Jeffereson Lab Hall A Coll.), 
Phys. Rev. C {\bf 64}, 038202 (2001).
%
\bibitem{Miller}
For recent discussions see, e.g., 
G.A. Miller, and M.R. Frank, Phys. Rev. C {\bf 65}, 065205 (2002).
%
\bibitem{Brodsky}
S.J. Brodsky and G.R. Farra, Phys. Rev. D {\bf 11}, 1309 (1975). 
%
\bibitem{Johann}
J. Haidenbauer, private communication.
%
\bibitem{PDG}
Particle Data Group, Phys. Rev. D {\bf 66}, (2002);
Eur. Phys. J. {\bf C3}, 1 (1998).
%
\bibitem{Benmer}
M. Benmerrouche, N. C. Mukhopadhyay, and J. F. Zhang,
Phys. Rev. D {\bf 51}, 3237 (1995);
J. F. Zhang, N. C. Mukhopadhyay and M. Benmerrouche,
Phys. Rev. C {\bf 52}, 1134 (1995).
%
\bibitem{Kanzo_inv}
K. Nakayama, J. Haidenbauer, C. Hanhart, and J. Speth, 
nucl-th/0302061.
%
%
%
%
\bibitem{MAINZ}
G. Blanpied et al., Phys. Rev. Lett. {\bf 79}, 4337 (1997);\\ 
R. Beck et al., Phys. Rev. C {\bf 61}, 035204 (2000).
%
\bibitem{Alex_omega}
G.I. Lykasov et al., Eur. Phys. J. A {\bf 6}, 71 (1999).
%
%
%
\bibitem{TN_inv}
K. Tsushima, and K. Nakayama, work in progress.
%
\bibitem{COSYd}
S. Barsov et al., nucl-ex/0305031
%
\bibitem{Cotanch}
S.R. Cotanch, and R.A. Williams, Phys. Lett. B {\bf 549}, 85 (2002). 
%
%
\bibitem{AlexK}
A. Sibirtsev, K. Tsushima, and A.W. Thomas, 
Phys. Lett. B {\bf 421}, 59 (1998).
%
\end{thebibliography}
\end{document}